\documentclass[aps,prl,epsfigure,twocolumn,superscriptaddress]{revtex4-1}
\usepackage[colorlinks=true,linkcolor=blue,urlcolor=blue,citecolor=blue,pdfusetitle]{hyperref}
\usepackage[utf8]{inputenc}
\usepackage[english]{babel}
\usepackage{amsmath}
\usepackage[caption = false]{subfig}
\usepackage{graphicx,epstopdf}
\usepackage{blindtext}
\usepackage[table,xcdraw]{xcolor}
\usepackage{lipsum}
\usepackage{amsfonts}
\usepackage{bbm}
\usepackage{amssymb}
\usepackage{enumerate}
\usepackage{color}
\usepackage{latexsym}
\usepackage{physics}
\usepackage{times,txfonts}

\newcommand{\Dcal}{\mathcal{D}}
\newcommand{\Ecal}{\mathcal{E}}
\newcommand{\Fcal}{\mathcal{F}}

\newcommand{\Mcal}{\mathcal{M}}

\newcommand{\Vcal}{\mathcal{V}}

\newcommand{\1}{\mathbbm{1}}

\newcommand{\interpro}[2]{\langle #1 | #2 \rangle}

\definecolor{greendark}{rgb}{0.0, 0.5, 0.0}

\begin{document}

\title{Quantum battery based on quantum discord at room temperature}

\author{C. Cruz}
\email{clebson.cruz@ufob.edu.br}
\affiliation{Grupo de Informa\c{c}\~{a}o Qu\^{a}ntica, Centro de Ci\^{e}ncias Exatas e das Tecnologias, Universidade Federal do Oeste da Bahia - Campus Reitor Edgard Santos. Rua Bertioga, 892, Morada Nobre I, 47810-059 Barreiras, Bahia, Brasil.}
\author{M. F. Anka}
\email{maronanka@id.uff.br}
\affiliation{Instituto de F\'{i}sica, Universidade Federal Fluminense, Av. Gal. Milton Tavares de Souza s/n, 24210-346 Niter\'{o}i, Rio de Janeiro, Brasil.}
\author{M. S. Reis}
\email{marioreis@id.uff.br}
\affiliation{Instituto de F\'{i}sica, Universidade Federal Fluminense, Av. Gal. Milton Tavares de Souza s/n, 24210-346 Niter\'{o}i, Rio de Janeiro, Brasil.}

\author{R. Bachelard}
\email{romain@df.ufscar.br}
\affiliation{Departamento de Física, Universidade Federal de São Carlos, Rodovia Washington Luís, km 235 - SP-310, 13565-905 São Carlos, SP, Brasil}

\author{Alan C. Santos}
\email{ac\_santos@df.ufscar.br}
\affiliation{Departamento de Física, Universidade Federal de São Carlos, Rodovia Washington Luís, km 235 - SP-310, 13565-905 São Carlos, SP, Brasil}

%\date{\today}

\begin{abstract}
	{The study of advanced quantum devices for energy storage has attracted the attention of the scientific community in the past few years. Although several theoretical progresses have been achieved recently, experimental proposals of platforms operating as quantum batteries under ambient conditions are still lacking. In this context, this work presents a feasible realization of a quantum battery in a carboxylate-based metal complex, which can store a finite amount of extractable work under the form of quantum discord at room temperature, and recharge by thermalization with a reservoir. Moreover, the stored work can be evaluated through non-destructive measurements of the compound's magnetic susceptibility. These results pave the way for the development of enhanced energy storage platforms through material engineering.}
\end{abstract}
\maketitle

\textbf{Introduction.}  
Batteries are common components in many technological devices, storing  different types of energy and converting it under the form of electric current
\cite{li201830}. Over the past years, the quest for an efficient use of energy has boosted  the development of reliable mechanisms for energy storage~\cite{Liu:19,manzano2013scientific}. Recently, a novel class of batteries has attracted the attention of the scientific community, namely,  quantum batteries (QBs). Chemical batteries convert chemical energy into electric one through  reactions between two species with different chemical properties~\cite{li201830}. Differently, QBs are constituted of quantum systems, and exploit the superposition principle of states and its quantum correlations. Based on these, QBs are able to store an amount of work from the quantum states,  called \textit{ergotropy}, which is extracted to power quantum devices~\cite{Ferraro:18,Alicki:13,Binder:15,Santos:20c,CampbellBatteries,PRL2017Binder,PRL_Andolina,Baris:20,Santos:19-a,PhysRevE.102.042111,PhysRevLett.125.180603,Le:18}-not necessarily as electric power.

However, there are several challenges to overcome before QBs are put into practical use, such as storing ergotropy at room temperature (i.e., preserving quantum correlations or coherence), offering a non-destructive access to the amount of stored work, and a charge lifetime comparable with that of the conventional classical batteries. Recently, several devices have been proposed as QBs, based on different quantum systems, for example, using solid-state and quantum optical systems, such as superconducting devices \cite{strambini2020josephson}, circuit-QED~\cite{PRL_Andolina,Rossini:20}, and two-level emitters into waveguides~\cite{Alexia:20}. However, the platforms proposed so far for QBs hardly fulfill the above requirements, and the possibility to place QBs into practical use is still under debate~\cite{Alexia:20,strambini2020josephson}. 

\begin{figure}[t!]
	\centering
	\includegraphics[scale=0.395]{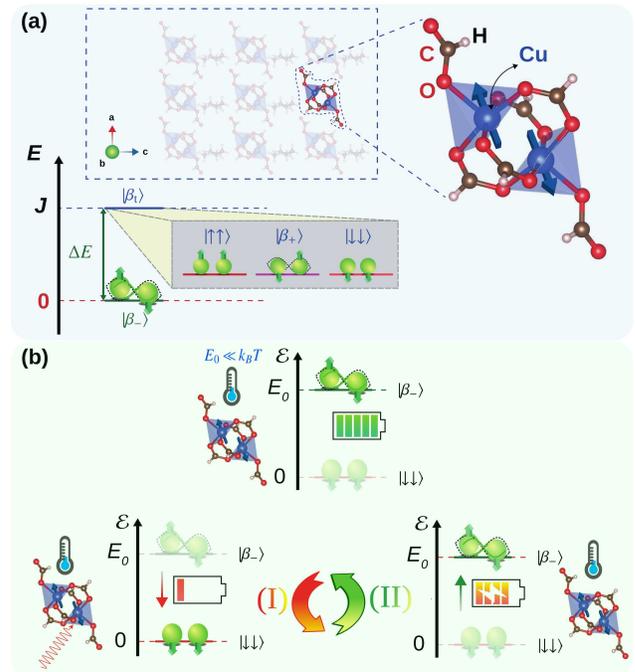}
	\caption{{\color{blue}(a)} Sketch of a Cu$_2$(HCOO)$_4$(HCOOH)$_2$(C$_4$H$_{10}$N$_2$) low-dimensional metal complex \cite{ccdc,vesta}, with a polyhedron representation of the molecule. The internal energy levels of the compound composed by a singlet ground-state $\ket{\beta_{-}}$ and a triplet degenerate subspace ($\ket{\beta_{t}}$). 
		The arrows stand for the effective spins stemming from the magnetic structure. {\color{blue}(b)} Sketch of the discharging and charging  processes. (top) The material in thermal equilibrium with a reservoir at temperature $T$ (with maximum energy stored achieved in 83 K) is in its singlet ground state, with energy stored $E_0$: (bottom) (I) an external stimulus (red arrow) drives it to the state $\ket{\downarrow\downarrow}$ (with zero energy stored) by consuming the stored ergotropy; (II) removing the stimulus charges the QB, by increasing the population of $\ket{\beta_{-}}$, bringing it back to the singlet ground-state.}
	\label{structure}
\end{figure}

In this context, low-dimensional metal complexes (LDMC), such as quantum antiferromagnets~\cite{cruz,souza,mario2,kova2020unconventional,cruz2020quantifying,souza2,he2017quantum,he2017quantum,breunig2017quantum,kova2020unconventional}, appear as promising platforms to implement QBs prototypes. Indeed, these complexes present a molecular structure that shields them from environment fluctuations. This includes fluctuations in temperature~\cite{mario2,kova2020unconventional}, magnetic field~\cite{kova2020unconventional,cruz2020quantifying,souza2}, and pressure~\cite{cruz2017influence,cruz2020quantifying}. LDMCs effectively behave as a two-qubit system (see Fig.\ref{structure}(a)), which can hold stable quantum correlations above room temperature \cite{cruz,mario2,souza}.

In this work, we propose metal complexes as room-temperature operating QBs, where the working substance is a dinuclear copper (II) carboxylate-based metal-organic complex~\cite{cruz}. The ergotropy is stored in the quantum discord between the spins, and the available ergotropy can be measured non-destructively by monitoring the magnetic susceptibility of the compound.
Finally, the carboxylate structure proves robust against self-discharging processes~\cite{Santos:20d}. 

\textbf{Metal complexes as effective two-qubit systems.} In this work, we focus on a metal-organic dinuclear cooper (II) compound with chemical formula Cu$_2$(HCOO)$_4$(HCOOH)$_2$(C$_4$H$_{10}$N$_2$), whose crystalline structure is obtained through single-crystal X-ray analysis \cite{ccdc} (see Fig.~\ref{structure}{\color{blue}a}). The system is prepared using dehydrated copper (II) nitrate (Cu(NO$_3$)$_2$), piperazine (C$_4$H$_{10}$N$_2$), formic acid, ethanol and distilled water; and refluxed at $~425$ K for 3~hours. The piperazine organic compound yields a porous environment which allows for the production of metal-organic frameworks with magnetically isolated dimers of Cu(II). 

The reduced magnetic unit is then formed by two metallic centers of Cu(II), with electronic configuration $d^{9}$ and $s=1/2$, in a dimeric  tetraformate unit. The \textit{syn-syn} bond between them, characteristic of carboxylate-based metal complex, yields a short intermolecular separation of $2.68(2)$~$\AA$~\cite{cruz,souza}. This provides a nearly-ideal realization of an isolated two-qubit system~\cite{yurishchev2011quantum,gaita2019molecular,moreno2018molecular}. It is worth noting that, despite the specificity of the present LDMC, this behavior is actually encountered in a broad range of metal complexes, thanks to their large intramolecular interaction energy, as compared to the intermolecular ones~\cite{cruz2020quantifying,yurishchev2011quantum,souza,souza2,kova2020unconventional,chakraborty2019magnetocaloric,yurishchev2011quantum,gaita2019molecular,moreno2018molecular}. The energy of the two coupled qubit system is provided by the following Hamiltonian:
\begin{align}
	H = E_{0} \left(S_{1}^{(z)}   + S_{2} ^{(z)}\right)+J \left( \vec{S}_{1} \cdot \vec{S}_{2} \right).
\end{align}
Above, $S_{n}^{(k)}\!=\!(\hbar/2)\sigma_{n}^{k}$, where $\sigma_{n}^{k}$ represent the Pauli matrices and $k\!\in\!\{x,y,z\}$. The first right-hand term is the Zeeman Hamiltonian~\cite{mario}, where $E_{0}\!=\!\mu_B g_z B_z$, $g_z$ the isotropic Land\'{e} factor; $\mu_B$ represents the Bohr magneton; and $B_z$ is the external (fixed) magnetic field intensity. Such term describes the energy levels of each cell of the battery (qubit): $H_{0}\ket{\epsilon_{i}}\!=\!\epsilon_{i}\ket{\epsilon_{i}}$, with $\epsilon_i\!\propto\!E_{0}$ and $\ket{\epsilon_{i}}$ the eigenvalue and eigenstate, respectively. In particular, the energy scale $E_0$ (and, as we shall see later, the ergotropy of the system - Fig.~\ref{structure}{\color{blue}b}), is proportional to the fixed magnetic field $B_z$. The Heisenberg Hamiltonian $H_{\text{int}}= J \left( \vec{S}_{1} \cdot \vec{S}_{2} \right)$ corresponds to the internal interaction between Cu(II) ions (intra-cell interaction), while $J$ represents the magnetic coupling constant.

The \textit{syn-syn} metal carboxylate conformation yields a very short metal-to-metal magnetic interaction, leading to a huge magnetic coupling $J/k_{B}\!=\!748$ K between Cu(II) cells. This allows for the existence of stable quantum correlations above room temperature~\cite{mario2,souza}. Consequently, as sketched in Fig.~\ref{structure}{\color{blue}a}, the energy levels of the system is composed by the singlet state $\ket{\beta_{-}}$ with energy $E_{-} = 0$ and the triply-degenerate subspace $\{\ket{\beta_{\text{t}}}\}\!=\!\{\ket{\beta_{+}}, \ket{\downarrow\downarrow}, \ket{\uparrow\uparrow}\}$, with energy $E_{\text{t}} = J$, where we have defined $\ket{\beta_{\pm}}\!=\!(\ket{\downarrow\uparrow}\pm\ket{\uparrow\downarrow})/\sqrt{2}$. Thus, the gap $\Delta E > 0$ between the ground and first excited ground state indicates that the system will be in a entangled singlet ground state with anti-parallel alignment (quantum antiferromagnet). 

In particular, for this class of materials, the static magnetic field $B_z$ splits the energy levels of the compound, inducing a quantum level crossing~\cite{chakraborty2019magnetocaloric,cruz2020quantifying,breunig2017quantum}, changing its corresponding populations \cite{cruz2020quantifying} (for details, see ref.~\cite{SupInf}). However, the strong magnetic coupling, yielded by the \textit{syn-syn} bond between the Cu(II) cells, leads to a crossing field of $B_c\sim\!556$~T. This very large value of the crossing field is due to the large gap ($J_{\text{int}}/k_{B}\!=\!748$ K) between the ground (singlet entangled state) and the first excited states (triplet separable state) yielded by the \textit{syn-syn} bond between the Cu(II) cells (see Fig.~\ref{structure}{\color{blue}a}). Therefore, for any $B_{z}\ll B_{c}$ ($E_{0}\ll J$), the system behaves as an effective two-level one, with cycle between the singlet ground state $\ket{\beta_{-}}$ and the excited triplet $\ket{\beta_{\text{t}}}$ one. 

The density matrix for the coupled system in thermal equilibrium can be written, in the energy basis, as an X-state:
\begin{align}
	\rho(T,B_z) &=&\frac{e^{{\zeta}}}{2Z}\left[
	\begin{matrix} 2e^{\beta E_{0}} & 0 & 0 & 0 \\
		0& {1 +  e^{-4{\zeta}}} & {1 -  e^{-4{\zeta}}}  & 0\\
		0& {1 -  e^{-4{\zeta}}}  & {1 +  e^{-4{\zeta}}}  & 0\\
		0& 0& 0 & 2e^{-\beta E_{0}} 
	\end{matrix} \right]~,
	\label{longitudinal}
\end{align}
where $Z(T,B_z)\!=\!e^{{\zeta}} + e^{-3{\zeta}} + 2 e^{{\zeta}} \cosh\left(\beta E_{0}\right)$ is the partition function, $\beta\!=\!1/k_BT$ and ${\zeta}\!=\!\beta J/4$. 

\textbf{Extractable work (ergotropy).} 
Unlike classical batteries, a QB is characterized by a finite amount of work in a quantum system, which can be extracted via unitary process~\cite{Allahverdyan:04}. The maximum amount of available work extractable via unitary processes is called \textit{ergotropy}~\cite{Liu:19,Alicki:13,PhysRevE.102.042111}.

Given the system state $\rho(T,B_z)$ and the QB internal spectrum of $H_{0}$, living in an $N$-dimensional Hilbert space, the ergotropy is defined as the amount of work which can be extracted by unitary operations $V$:
\begin{equation}
	\Ecal(T,B_z)\!=\!\mbox{Tr}\left[\rho(T,B_z) H_{0}\right] - \min_{V\in \Vcal}\left\{\mbox{Tr}\left[V\rho(T,B_z) V^\dagger H_{0}\right]\right\},
	\label{Ergotropy}
\end{equation}
where the minimization is taken over the set $\Vcal$ of all unitary operators acting on the system~\cite{Allahverdyan:04}. The energy eigenvalues of the battery self-Hamiltonian $H_0$ are $\epsilon_{1}\!\leq\!\epsilon_{2}\!\leq\!\cdots\!\leq\!\epsilon_{N}$; and the eigenvalues of $\rho(T,B_z)$ are $\varrho_{1}\!\geq\!\varrho_{2}\!\geq\!\cdots\!\geq\!\varrho_{N}$, associated to eigenvectors $\ket{\varrho_{n}}$~\cite{SupInf}. Then, the ergotropy in the Eq.~\eqref{Ergotropy} can be written as~\cite{Allahverdyan:04}:
\begin{align}
	\Ecal(T,B_z) = \sum\nolimits_{i,n}^{N,N} \varrho_{n}(T,B_z) \epsilon_{i} \left( |\interpro{\varrho_{n}}{\epsilon_{i}}|^2 - \delta_{ni} \right) , \label{ErgotropyXstates}
\end{align}	
where we note that this definition is tied to a specific ordering of the eigenvalues $\varrho_{n}$ of Eq.~\eqref{longitudinal} and $H_0$, due to the $\delta_{ni}$ term. This leads to the following expression for the {\it ergtropy per molecule}, in thermal equilibrium:
\begin{equation}
	\Ecal_{\text{Cu(II)}} = \
	E_{0} \left\{\frac{1 - e^{{4\zeta}}\left[\cosh \left(\beta E_{0}\right)-3 \sinh \left(\beta E_{0}\right)\right]}{e^{{4\zeta}}\left[2 \cosh \left(\beta E_{0}\right)+1\right]+1}\right\}.
	\label{ErgotropyCarboxylate}
\end{equation}

On the other hand, the enormous gap between the ground (singlet - entangled) and the first excited (triplet - separable) state provided by the \textit{syn-syn} bound, allows us to address the system in the regime of magnetic susceptibility at $E_0\ll k_{B}T$. From Eq.~\eqref{ErgotropyCarboxylate}, it is possible to write the ergotropy $\Ecal_{\text{Cu(II)}}$, in terms of the Bleaney-Bowers magnetic susceptibility equation  ~\cite{bleaney1952anomalous} (for details, see ref.~\cite{SupInf})
\begin{equation}
	\Ecal_{E_0\ll k_{B}T}(T)  = \
	E_{0} \frac{k_BT\chi(T)}{4N_Ag^2\mu_B^2}\left[ e^{-4\zeta} -1\right]~, 
	\label{Ergotropysus}
\end{equation}
where $N_A$ is the Avogadro number. As detailed in~\cite{SupInf}, this magnetic susceptibility regime is valid in the limit of small magnetic  fields $B_{z}\ll B_{c}$ compared to the crossing one ($B_c\!\sim\!556$~T) \cite{chakraborty2019magnetocaloric}.

Therefore, Eq.~\eqref{Ergotropysus} represents one of the main results of this letter, in which the evaluation of the amount of stored ergotropy in metal complexes are experimentally accessible, by measuring a macroscopic property of the system: the magnetic susceptibility.

\begin{figure}[t!]
	\centering
	\subfloat[]{\includegraphics[scale=0.35]{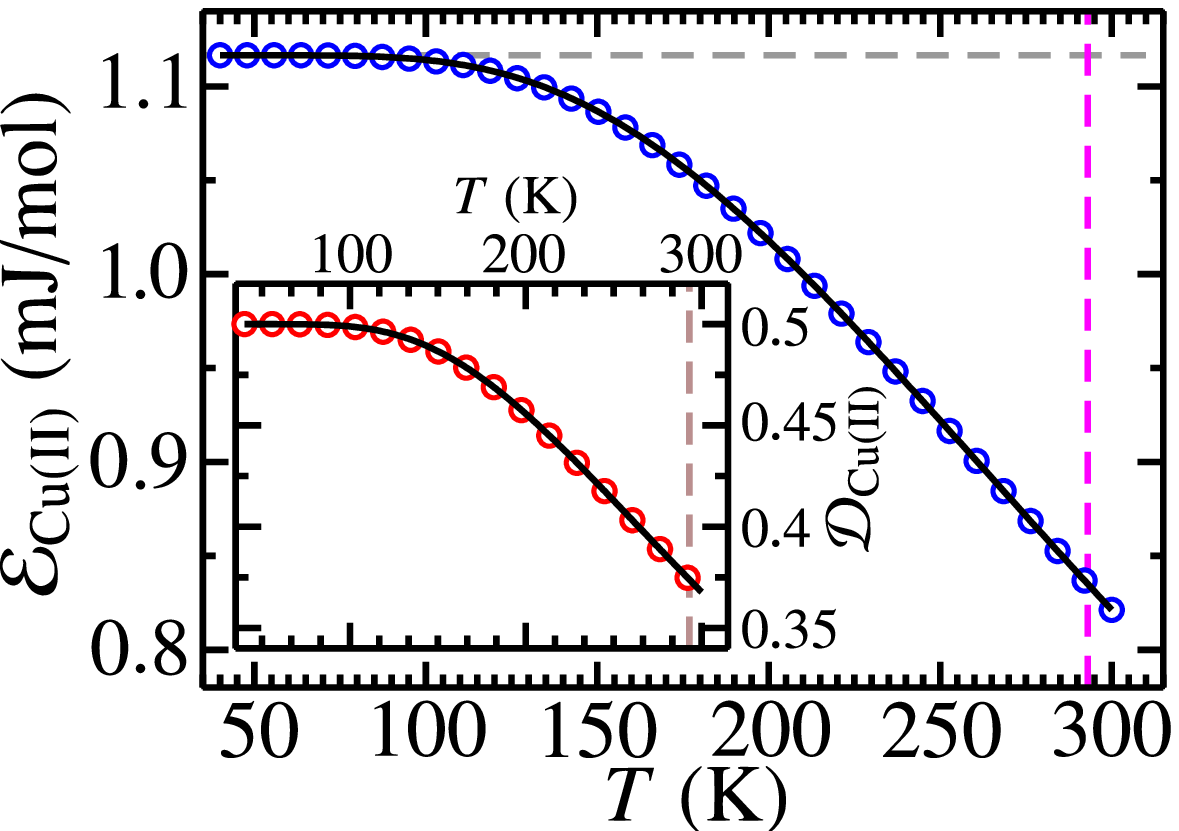}\label{Fig2a}}
	\subfloat[]{\includegraphics[scale=0.35]{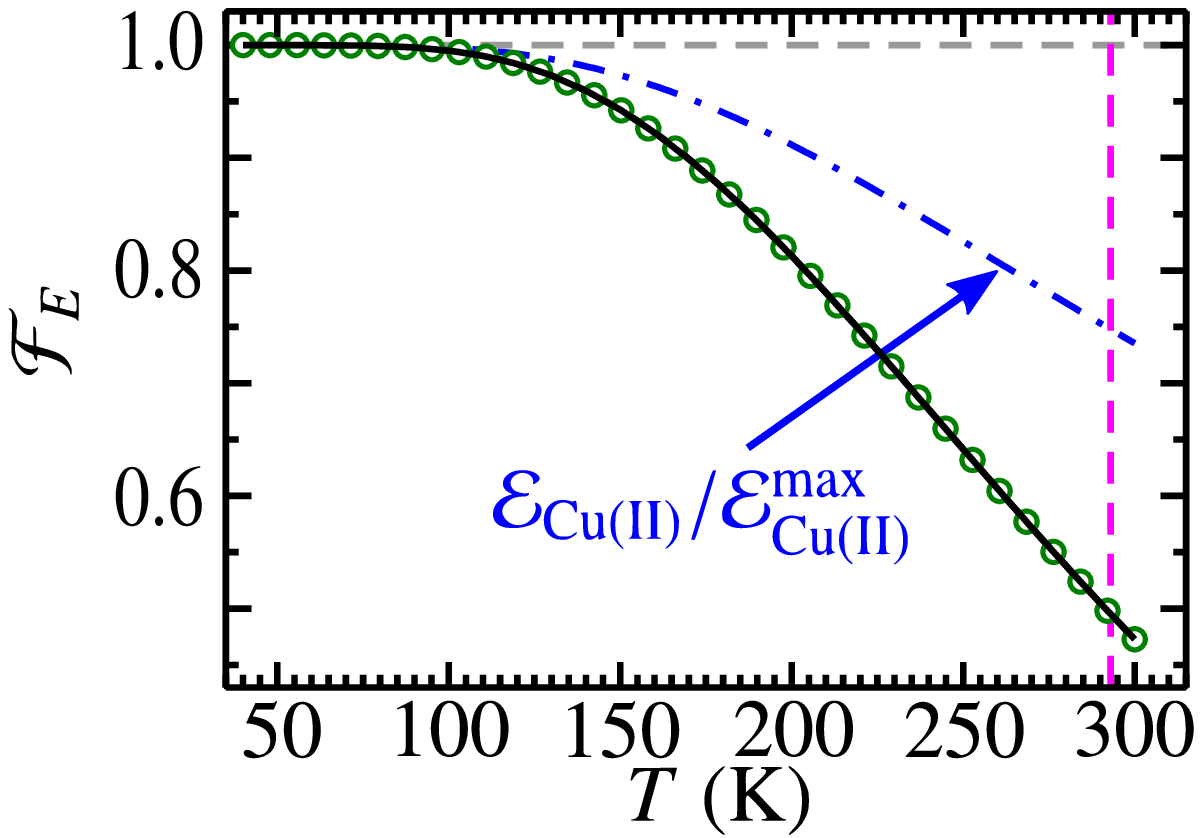}\label{Fig2b}}
	\caption{{\color{blue}(a)} Experimental measurement of ergotropy, in mili Joule per mole, as function of the environment temperature $T$ (in Kelvin) for the metal-carboxylate QB proposed. The inset shows the experimental quantum discord  as function of the temperature. Black solid line denote theoretical results and open circles show the experimental result, computed for the Earth's magnetic field $B_z\!=\!10^{-4}$~T. Vertical dashed line indicates room temperature $T\!=\!293$~K. {\color{blue}(b)} Theoretical (solid line curve) and experimental result (open circles) for entanglement of formation, compared with the theoretical result for normalized ergotropy (dot-dashed line).}
	\label{QuantitiesFig}
\end{figure}

\textbf{Ergotropy measurement and quantumness of metal-carboxylate QBs.} Up to date, proposals of QBs have been done through an ergotropy destructive readout, through state tomography \cite{Liu:19,Ferraro:18,Le:18,strambini2020josephson,Alexia:20}. The QB proposed here provides a different approach to read the stored ergotropy, by measuring the available work of a carboxylate-based QB accessible without altering the stored energy, this is a strong request for designing realistic QBs. Fig.~\ref{Fig2a} presents the theoretical (black line) and experimental (open circles) behavior of the ergotropy as a function of the temperature, where the experimental data was obtained from the measurement of the magnetic susceptibility of the substance. The theoretical curve was plotted by Eq.~\eqref{ErgotropyCarboxylate}, considering the Earth's magnetic field of $B_z\!=\!10^{-4}$~T, and the experimental parameters $g=2$ ($d^{9}$ ions) and $J/k_{B}\!=\!748$~K, in a good agreement with the laboratory environment in which batteries should operate. It is worth highlighting that by decreasing the temperature we get the maximum amount of ergotropy $\Ecal^{\text{max}}_{\text{Cu(II)}}\!\approx\!1.12$~mJ/mol, where values larger than $99.75\%$ of $\Ecal^{\text{max}}_{\text{Cu(II)}}$ can be reached for temperatures $T_{\text{max}}\!\leq\!100$~K reaching the $100\%$ below $83$ K. Moreover, as can be seen, at room temperature ($\sim293$~K), the amount of stored energy is $75\%$ of the maximum one. This suggests that the carboxylate-based QB could operate under daily life conditions.

To explore the metal-carboxylate as a QB, we now present experimental results that give access to the ergotropy, which here corresponds to the quantum discord. The characterization of the quantumness of the battery is done by computing entanglement of formation ($\Fcal_{E}$)~\cite{Hill:97,Wootters:98} and the quantum discord based on Schatten 1-norm $\Dcal(T)$~\cite{Ciccarello:14}, in terms of the magnetic susceptibility~\cite{SupInf}. First, we find that the ergotropy as function of the temperature is given by the Schatten 1-norm quantum discord as~\cite{SupInf}
\begin{equation}
	\Ecal_{E_0\ll k_{B}T}(T)  = 2E_0 \Dcal(T) , ~~ \forall ~~ J>0 .
	\label{discord}
\end{equation}

From above equation we observe that, for any dinuclear metal complex of spin-1/2 with an antiparallel alignment ($J\!>\!0$), the maximum amount of extracted work is related to the existence of genuinely quantum correlations beyond entanglement between the Cu(II) ions, which is quantified by quantum discord. The inset in Fig.~\ref{Fig2a} show experimental result for the discord, showing that the decay of the ergotropy with temperature corresponds to the decay of Schatten 1-norm quantum discord.

To see how the ergotropy is stored as quantum correlations beyond entanglement, we also consider the entanglement of formation $\Fcal_{E}$ as shown in Fig.~\ref{Fig2b}. To this end, we normalized the ergotropy by its maximum value $\Ecal^{\text{norm}}_{\text{Cu(II)}}\!=\!\Ecal_{\text{Cu(II)}}/\Ecal^{\text{max}}_{\text{Cu(II)}}$, and thus both quantities $\Fcal_{E}$ and $\Ecal^{\text{norm}}_{\text{Cu(II)}}$ are defined in the same interval (\textit{i.e.}, $[0,1]$). Fig.~\ref{Fig2b} suggests that the entanglement is not adequate to explain the qualitative behavior of the ergotropy as function of the temperature. Thus, as expected from Eq.~\eqref{discord}, there is ergotropy even without entanglement, but not without quantum discord. In addition, as theoretically detailed in~\cite{SupInf}, by increasing the range of temperature, one finds that above the temperature of entanglement $T_{\text{e}}=J /(\ln (3)k_{B})$ in which $\Fcal_{E}(T\!\geq\! T_{\text{e}})\!=\!0$, the ergotropy is is $16.6\%$ of the maximum one ($\Ecal^{\text{norm}}_{\text{Cu(II)}}(T\!\geq\!T_{\text{e}})\!>\!0$) and satisfies the Eq.~\eqref{discord}.
Therefore, these results lead to the conclusion that other quantum correlations beyond entanglement needs to taken into account to describe the amount of ergotropy stored in metal complexes. 

\textbf{Charging and discharging of metal-carboxylate QBs.} Due to the high magnetic interaction obtained from the \textit{syn-syn} bond in the metal-carboxylate compounds, the huge magnetic coupling ($J\!=\!748$ K) supports the existence of stable quantum correlations at room temperature ($T\approx293$~K) \cite{cruz}. More specifically, it allows the studied carboxylate-based metal complex to store a finite amount of ergotropy as quantum discord up to a threshold temperature $513$~K (obtained through thermogravimetric analysis reported in ref.~\cite{cruz}) under which this material starts to degrade. Moreover, the strong magnetic coupling makes the system practically immune to the magnetic field variations in the experimentally feasible region of small magnetic fields $B_z\!\ll\!B_c$. Hence, in this regime the role of magnetic field is to define the energy scale ($E_0$) of the battery, as it can be seen in Eq.~\eqref{Ergotropysus}. Therefore, this system does not need a time-dependent external field for the charging process, differently from the several kinds of QBs already proposed in literature~\cite{PRL2017Binder,PRL_Andolina,Santos:20c,Le:18,Kwon:18,Alexia:20,Kamin:20-2}. 

Since there is no additional decoherence channel, the dimeric cells are magnetically isolated, and the material does not degrade below the threshold temperature. Thus, the system remains in the state given by Eq.~\eqref{longitudinal} and it is able to store a finite amount of energy quantum discord stable at room temperature. Thus, the charge lifetime is defined by the existence of non-zero genuine quantum correlations between the cells of the battery. Since the ergrotopy increases as we decrease the temperature, one can charge the QB the submitting the battery to thermal contact with a cold reservoir, under presence of a static reference magnetic field $B_z\ll B_c$. Therefore, the charging process of this carboxylate-based QB is given by a mechanism known in the literature as charging assisted by thermalization~\cite{Hovhannisyan:20}. 

In this regard, a charging and discharging steps of a major cycle can be presented  for this QB, as sketched in  Fig.~\ref{structure}{\color{blue}b}: (I) an external stimulus (e.g., electromagnetic field pulses \cite{moreno2018molecular,gaita2019molecular}, or pressure \cite{cruz2017influence,cruz2020quantifying}), lowers the degree of quantum discord of the system, drives it to the triplet state subspace (more specifically the state $\ket{\downarrow\downarrow}$),
discharging the battery by consuming the stored work \cite{PhysRevLett.125.180603,Alexia:20,Hovhannisyan:20,PhysRevE.102.042111}; (II) removing the stimulus, the material, in thermal equilibrium with a reservoir, returns to the singlet ground-state by increasing the population of $\ket{\beta_{-}}$, charging the battery (see top figure of the charging-discharging process). It is worth noting that these two steps belong to a major cycle and further discussions will be presented elsewhere.

\textbf{Conclusions.} In summary, this Letter shows that carboxylate-based magnetic systems are promising platforms for QBs. The available work is stored as quantum discord, i.e., genuine quantum correlations, at room temperature. Due to the \textit{syn-syn} bound a strong magnetic interaction emerges,leading to a large gap between the singlet ground state and the first excited one (see Fig.~\ref{structure}), which makes this material suitable for engineering room temperature QBs. In this sense, an external (fixed) magnetic field $B_z$ can be used to tune the total available work for the battery. In addition, the ergotropy measurement of the QB proposed here is done by non-destructive experimental techniques associated with the measurement of the compounds magnetic susceptibility, and no energy is lost during the readout of the available work. The strong intra-cell interactions also make this system largely immune to the self-discharging process, which allows it to store energy as long as there is quantum discord in the system at room temperature and beyond, up to threshold temperature (513 K) in which the material degrades. Moreover, at room temperature, the efficiency of this complex is approximately $75\%$ of the maximum one (obtained at $T\!<\!83$~K), in which quantum correlations appear, once again, as a harvestable resource of great interest for quantum technologies~\cite{Sapienza:19}.

Since a realistic implementation of a quantum battery has not yet been settled in an LDMC, the role of its chemical aspects when operating it as a QB remains to be studied. Nevertheless, the results presented in this letter open a broad avenue for research of metal complexes as candidate platforms for QBs and the development of enhanced energy storage platforms through material engineering. For example, quantum correlations in metal complexes can be handled by controlling external parameters, as structural factors and thermodynamic properties~\cite{wasielewski2020exploiting,kova2020unconventional,cruz2017influence,cruz2020quantifying}. Thus, different kinds of control of the gap between the ground and excited state in dinuclear spin-1/2 metal complexes QBs can be done through material engineering techniques~\cite{gaita2019molecular,moreno2018molecular,wasielewski2020exploiting}. In addition, quantum properties of solid-state systems are drastically affected by thermal decohering effects, hindering the development of feasible quantum batteries that operate at room temperature. This letter shows that the study on carboxylate-based materials can change this view and paves the way for the enhancement and stability of the charging and energy storage processes in QBs.

\begin{acknowledgments}
	The authors would like to thank P. Brand\~{a}o and A. M. dos Santos for the compound's data. A. C. Santos is supported by São Paulo Research Foundation (FAPESP) (Grant No 2019/22685-1). R.\,B. benefited from Grants from S\~ao Paulo Research Foundation (FAPESP, Grants No. 2018/15554-5 and 2019/22685-1). R. B. benefited from grants from the National Council for Scientific and Technological Development (CNPq, Grant Nos.\,313886/2020-2 and 409946/2018-4). M.S. Reis thanks FAPERJ for financial support. M.S. Reis belongs to the INCT of Refrigeração e Termofísica funding by the National Counsel of Technological
	and Scientific Development (CNPq), Grant No.\,404023/2019-3. This study was financed in part by the \textit{Coordena\c{c}\~{a}o de Aperfei\c{c}oamento de Pessoal de N\'{i}vel Superior - Brasil} (CAPES) - finance code 001.
\end{acknowledgments}

\newpage

\onecolumngrid

\begin{center}
	%%%%%%%%% ABSTRACT TITLE
	{\large{ {\bf Supplemental Material for: \\Quantum battery based on quantum discord at room temperature}}}
	%%%%%%%%% ABSTRACT AUTHORS
	\vskip0.5\baselineskip{C. Cruz,$^{1,{\color{blue}\ast}}$ M. F. Anka,$^{2,{\color{blue}\dagger}}$ M. S. Reis,$^{2,{\color{blue}\ddagger}}$ R. Bachelard,$^{3,{\color{blue}\S}}$ and Alan C. Santos$^{3,{\color{blue}\P}}$}
	%%%%%%%%% AFFILIATION^{1}
	\vskip0.5\baselineskip{\em$^{1}$Grupo de Informa\c{c}\~{a}o Qu\^{a}ntica, Centro de Ci\^{e}ncias Exatas e das Tecnologias, Universidade Federal do Oeste da Bahia - Campus Reitor Edgard Santos. Rua Bertioga, 892, Morada Nobre I, 47810-059 Barreiras, Bahia, Brasil.}
	{\em$^{2}$Instituto de F\'{i}sica, Universidade Federal Fluminense, Av. Gal. Milton Tavares de Souza s/n, 24210-346 Niter\'{o}i, Rio de Janeiro, Brasil.}
	{\em$^{3}$Departamento de Física, Universidade Federal de São Carlos, Rodovia Washington Luís, km 235 - SP-310, 13565-905 São Carlos, SP, Brasil.}
	%%%%%%%%% ElectronicAddress^{1}
	\vskip0.5\baselineskip{$^{\color{blue}\ast}$clebson.cruz@ufob.edu.br, $^{\color{blue}\dagger}$maronanka@id.uff.br, $^{\color{blue}\ddagger}$marioreis@id.uff.br\\ $^{\color{blue}\S}$bachelard.romain@gmail.com, $^{\color{blue}\P}$ac\_santos@df.ufscar.br}
\end{center}

\twocolumngrid

\appendix

%%%%%%%%%% Prefix a "S" to all equations, figures, tables and reset the counter %%%%%%%%%%
\setcounter{equation}{0}
\setcounter{figure}{0}
\setcounter{table}{0}

\renewcommand{\theequation}{S\arabic{equation}}
\renewcommand{\thefigure}{S\arabic{figure}}

\section{Quantum level-crossing in a metal complex}

\begin{figure}[!ht]
	\centering
	{\includegraphics[scale=0.37]{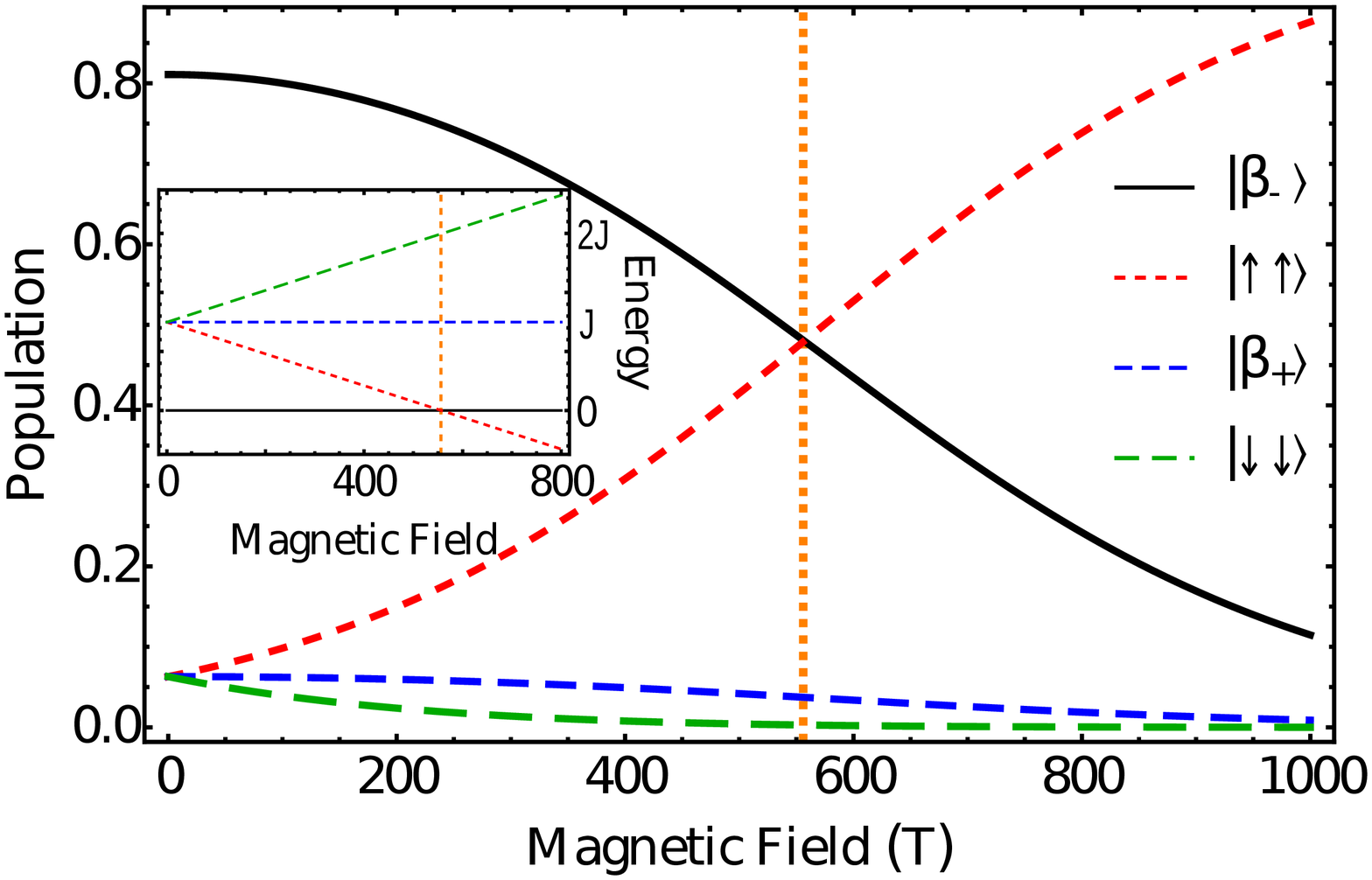}}
	\caption{(a) Populations, Eqs.~\eqref{x1}--\eqref{x4}, as a function of the magnetic field for different values of temperature.  The inset shows the the magnetic field dependence of the energy levels. Vertical (orange) dotted line represents the  value of the crossing field, as can be seen the population crossover occur at $B_c = 556$~T.}
	\label{population}
\end{figure}

Quantum antiferromagnets typically present a quantum level-crossing in the presence of an external magnetic field $B_z$, splitting its energy levels and changing the corresponding populations~\cite{chakraborty2019magnetocaloric,cruz2020quantifying,breunig2017quantum}. In particular for the carboxylate based metal complex described in the main text, the magnetic field splits the degeneracy of the triply-degenerate subspace $\{\ket{\beta_{\text{t}}}\}\!=\!\{\ket{\beta_{+}}, \ket{\downarrow\downarrow}, \ket{\uparrow\uparrow}\}$, inducing a level-crossing between its singlet-ground state and the first excited one when the field reaches a critical value \cite{chakraborty2019magnetocaloric,breunig2017quantum}. This  critical value can be calculated through the evaluation of the populations. From the Hamiltonian:
\begin{align}
H = E_{0} \left(S_{1}^{(z)}   + S_{2} ^{(z)}\right)+J \left( \vec{S}_{1} \cdot \vec{S}_{2} \right),
\label{hamiltonian}
\end{align}
and the spectral decomposition of the state $\rho(T,B_z)=e^{-\beta H}/Z$, where $Z=\mbox{Tr}{\left[e^{-\beta {H}}\right]}$, the battery state eigenvalues $\varrho_{i}$ (population) and its corresponding eigenvectors can be written as:
\begin{align}
\varrho_{1}&= \frac{1}{1 + e^{\zeta} ( 1 +  2 \cosh\left(\beta E_{0}\right))} \rightarrow \ket{\beta_{-}}, \label{x1}\\
\varrho_{2}&=\frac{ e^{\zeta + \beta E_{0}}}{1 + e^{\zeta} ( 1 +  2 \cosh\left(\beta E_{0}\right))} \rightarrow \ket{\uparrow\uparrow},\label{x2}\\
\varrho_{3}&=\frac{ e^{\zeta}}{1 + e^{\zeta} ( 1 +  2 \cosh\left(\beta E_{0}\right))} \rightarrow \ket{\beta_{+}},\label{x3}\\
\varrho_{4}&=\frac{e^{\zeta - \beta E_{0}}}{1 + e^{\zeta} ( 1 +  2 \cosh\left(\beta E_{0}\right))} \rightarrow \ket{\downarrow\downarrow}, \label{x4}
\end{align}
where $\ket{\beta_{\pm}}\!=\!(\ket{\downarrow\uparrow}\pm\ket{\uparrow\downarrow})/\sqrt{2}$; $\beta = 1/k_BT$; ${\zeta}\!=\! -\beta J$; and  $E_{0}\!=\!\mu_B g B_z$, with $g$ being the isotropic Land\'{e} factor and $\mu_B$ the Bohr magneton. Since the quantum level crossing is temperature independent, Fig. \ref{population} shows the populations $\varrho_{n}$ as function of the magnetic field at room temperature $T=293$~K, obtained from the experimental parameters $g=2$ ($d^{9}$ ions) and $J/k_{B}\!=\!748$~K~\cite{cruz}. The inset shows the magnetic field dependence of the energy levels
\begin{align}
E_{1}&= 0 \rightarrow \ket{\beta_{-}}, \label{e1}\\
E_{2}&=J - E_{0}, \rightarrow \ket{\uparrow\uparrow},\label{e2}\\
E_{3}&=J \rightarrow \ket{\beta_{+}},\label{e3}\\
E_{4}&=J+E_{0} \rightarrow \ket{\downarrow\downarrow}, \label{e4}
\end{align}
obtained from the spectral decomposition of the system Hamiltonian, Eq.~\eqref{hamiltonian}. As can be seem, the magnetic field over the populations  is driven our system from a maximally entangled ground state $\ket{\beta_{-}}\!=\!\left[\ket{\downarrow\uparrow} - \ket{\uparrow\downarrow} \right]/\sqrt{2}$, to a separable ground state $\ket{\uparrow\uparrow}$,  when the magnetic field surpass the critical value $B_c\approx 556$~T ($E_{0}\!=\! J$). This huge value of crossing field is a consequence of the strong magnetic coupling ($J/k_{B}\!=\!748$~K) yielded by the $syn-syn$ bound, which shields the compound from environment fluctuations~\cite{mario2,kova2020unconventional,cruz2020quantifying,souza2,cruz2017influence}. Thus, from Eq.~\eqref{e1}-\eqref{e4}, for any $B_{z}\ll B_{c}$ ($E_{0}\ll J$), the compound behaves as an effective two-level system. 

\section{Ergotropy of dinuclear spin-1/2 metal complexes}

Quantum systems can be defined as a quantum batteries when a finite amount of work, stored in its quantum states can be extracted via unitary process~\cite{Liu:19,Alicki:13,PhysRevE.102.042111}. Considering the carboxylate-based metal complex described by Eq.~\eqref{hamiltonian}, one can define the battery Hamiltonian as 
\begin{equation}
H = H_{0}+ H_{int},
\label{batery_hamiltonian}
\end{equation}
where $H_{0}= E_{0} \left(S_{1}^{(z)}+S_{2}^{(z)}\right)$ is the battery self-Hamiltonian, which defines the energy scale $\epsilon_{1}\!\leq\!\epsilon_{2}\!\leq\!\cdots\!\leq\!\epsilon_{N}$ and eigenstates $\ket{\epsilon_{i}}$ of the quantum battery, satisfying the eigenvalue equation $H_{0}\ket{\epsilon_{i}}\!=\!\epsilon_{i}\ket{\epsilon_{i}}$; and $H_{int}=J \left( \vec{S}_{1} \cdot \vec{S}_{2} \right)$ defines an internal interaction of the system. The system has a corresponding density matrix, defined in thermal equilibrium as:
\begin{align}
\rho(T,B_z) &=&\frac{e^{{\zeta}}}{2Z}\left[
\begin{matrix} 2e^{\beta E_{0}} & 0 & 0 & 0 \\
0& {1 +  e^{-4{\zeta}}} & {1 -  e^{-4{\zeta}}}  & 0\\
0& {1 -  e^{-4{\zeta}}}  & {1 +  e^{-4{\zeta}}}  & 0\\
0& 0& 0 & 2e^{-\beta E_{0}} 
\end{matrix} \right]~,
\label{Sup-longitudinal}
\end{align}
where $Z(T,B_z)\!=\!e^{{\zeta}} + e^{-3{\zeta}} + 2 e^{{\zeta}} \cosh\left(\beta E_{0}\right)$ is the partition function. 

For non-pure states ($Tr\{\rho^2\}\!\neq\!1$), the amount of available work which can be extracted from unitary operations $V$ is not given by the internal energy of the system $U\!=\!Tr\{H_{0}\rho\}$, but rather by the maximum amount of available work extractable via unitary processes, defined by the \textit{ergotropy}~\cite{Allahverdyan:04,Liu:19,Alicki:13,PhysRevE.102.042111}:
\begin{equation}
\Ecal(T,B_z)\!=\!\mbox{Tr}\left[\rho(T,B_z) H_{0}\right] - \min_{V\in \Vcal}\left\{\mbox{Tr}\left[V\rho(T,B_z) V^\dagger H_{0}\right]\right\},
\label{Sup-Ergotropy}
\end{equation}
where the minimization, taken over the set $\Vcal$ of all unitary operators which acts on the system~\cite{Allahverdyan:04} is obtained by writing Eq.~\eqref{Sup-longitudinal} in its spectral decomposition, with the populations $\varrho_{n}$, Eq.~\eqref{e1}-\eqref{e4}, in decreasing order $\varrho_{1}\!\geq\!\varrho_{2}\!\geq\!\cdots\!\geq\!\varrho_{N}$, and the eigenvalues of the self-Hamiltonian in the increasing order $\epsilon_{1}\!\leq\!\epsilon_{2}\!\leq\!\cdots\!\leq\!\epsilon_{N}$. Then, the maximum amount of available work which can be extracted from a quantum state, is given by  ~\cite{Allahverdyan:04,Gianluca:17,Alicki:13}:
\begin{align}
\Ecal = \sum\nolimits_{i,n}^{N,N} \varrho_{n} \epsilon_{i} \left( |\interpro{\varrho_{n}}{\epsilon_{i}}|^2 - \delta_{ni} \right). \label{Sup-ErgotropyXstates}
\end{align}

In particular, as mentioned before, due to the characteristics quantum level crossing presented by quantum antiferromagnets \cite{chakraborty2019magnetocaloric,cruz2020quantifying,breunig2017quantum}, it is  possible to establish two orderings of the quantities $\varrho_{n}$. Consequently, for dinuclear spin-1/2 metal complexes, the ergotropy can defined in two physical regimes: one for $E_{0}\!<\! J$, corresponding to $B_{z}<B_{c}$; and other for $E_{0}\!\geq\! J$, corresponding to $B_{z}\geq B_{c}$. Thus, from Eq.~(\ref{Sup-ErgotropyXstates}), we obtain the \textit{ergotropy per molecule} in thermal equilibrium with a reservoir at temperature $T$ under presence of a static reference magnetic field $B_{z}$ as
\begin{equation}
\Ecal= \left\{
\begin{aligned}
E_{0} \left\{\frac{1 - e^{{\zeta}}\left[\cosh \left(\beta E_{0}\right)-3 \sinh \left(\beta E_{0}\right)\right]}{e^{{\zeta}}\left[2 \cosh \left(\beta E_{0}\right)+1\right]+1}\right\}, ~~ E_{0}< {J} \\
\frac{4 E_{0}e^{{\zeta}}\sinh \left(\beta E_{0}\right)}{e^{{\zeta}}\left(2 \cosh \left(\beta E_{0}\right)+1\right)+1}, ~~ E_{0}\geq {J}
\end{aligned}
\right.~.
\label{Sup-ErgotropyCarboxylate}
\end{equation}

In particular, for the carboxylate-based metal complex described in the main text, the regime $B_{z}\geq B_{c}$ are not accessible by usual experimental techniques, due to the huge value of crossing field ($556$~T), consequence of the \textit{syn-syn} bound. Therefore, the experimentally  feasible available work, stored in the metal-carboxylate compound, can be found only in the regime $E_{0}\!<\! J_{\text{int}}$.

\section{Ergotropy and magnetic susceptibility}

From the system Hamiltonian, Eq.~\eqref{hamiltonian}, we obtain the canonical partition function $Z(T,B_z)\!=\!e^{{\zeta}} + e^{-3{\zeta}} + 2 e^{{\zeta}} \cosh\left(\beta E_{0}\right)$. Then, one can obtain the magnetization of a dinuclear spin-1/2 metal complex as~\cite{mario}
\begin{align}
\Mcal_z(T,B_z) &=  \frac{Nk_{B}T}{Z (T,B_z)}\frac{\partial}{\partial B_z} Z(T,B_z)~, \nonumber \\
&= \frac{Ng\mu_B\left( e^{2\beta E_0} - 1\right)}{ 1 +  e^{\beta E_0} \left( 1+ e^{-\zeta} + e^{\beta E_0}\right)}~.
\label{magnetization}
\end{align}

\begin{figure}[!ht]
	\centering
	\includegraphics[scale=0.5]{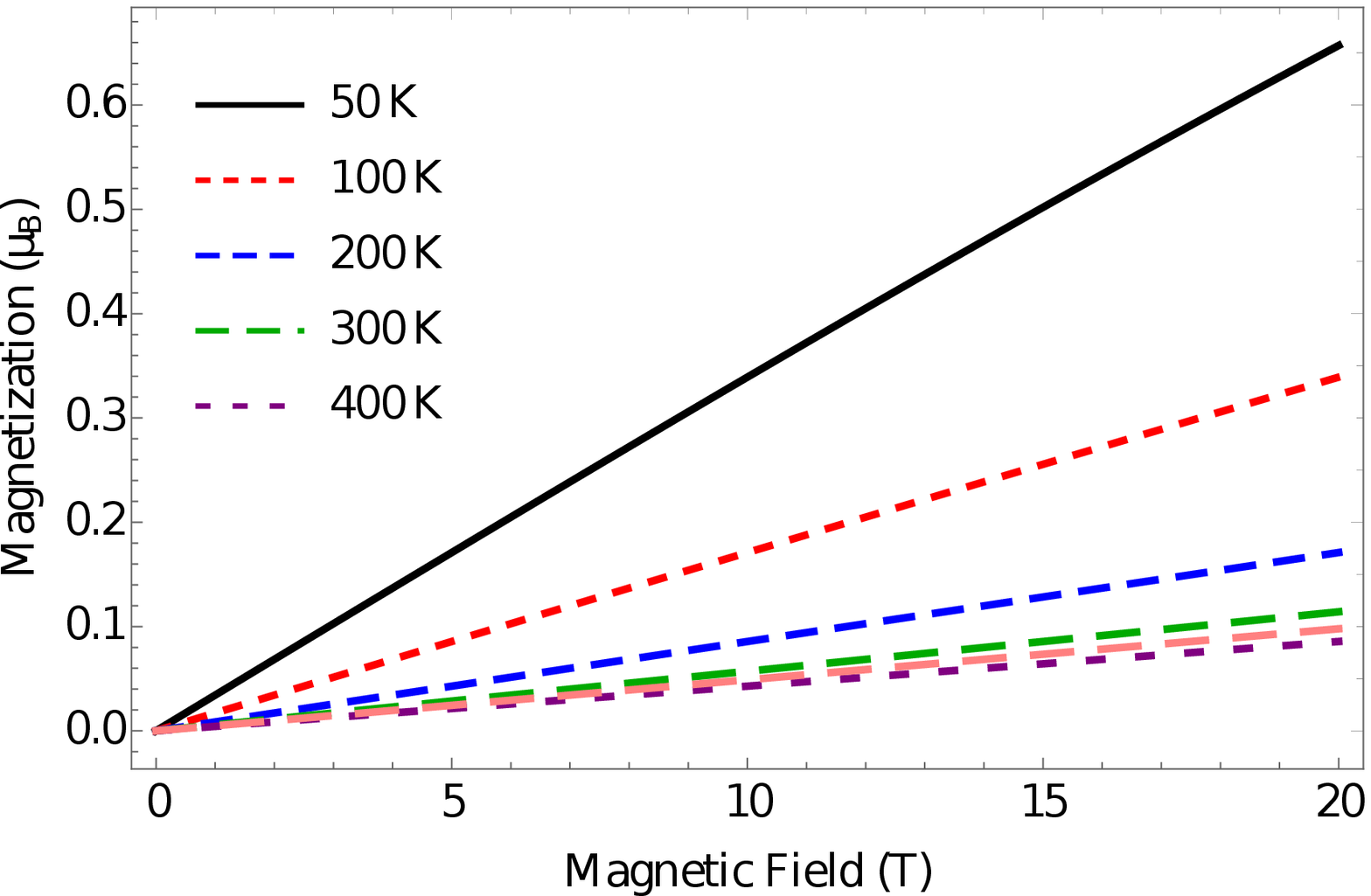}
	\caption{Magnetization $\Mcal_z$ as a function of the magnetic field $B_z$ for the temperature range under which the magnetic susceptibility of the described metal complex is measured. As can be seen, in the entire temperature range the system remains in the regime of magnetic susceptibility (linear magnetization) up to dozens of Tesla.}
	\label{mag}
\end{figure}

Thus, the regime of magnetic susceptibility of this system is reached under low values of magnetic field in which the magnetization has the linear dependence $\Mcal_z  = B_z \chi$ \cite{mario}. In particular, for the described carboxylate-based metal complex, with the experimental parameters $J/k_{B}\!=\! 748$ K  and $g = 2$ \cite{cruz}, we obtain the theoretical magnetization $\Mcal_z(T,B_z)$ as a function of the magnetic field for different temperatures (Fig. \ref{mag}). As can be seen in Fig. \ref{mag}, for the temperature regime under which the magnetic susceptibility is measured, the system remains in the limit of small magnetic fields even for values up to dozens of Teslas. In a general way, the enormous gap between the ground (singlet - entangled) and the first excited (triplet - separable) state provided by the \textit{syn-syn} bound, allows us to address the system in the regime of magnetic susceptibility at $E_0\ll k_{B}T$ when $B_z\ll B_c$.

Thus, applying the limit $E_0\ll k_{B}T$ in Eq. (\ref{magnetization}) we can obtain the the Bleaney-Bowers equation for magnetic susceptibility for two spin-$1/2$ interacting systems~\cite{bleaney1952anomalous}
\begin{align}
\chi(T) &=  \lim_{E_0\ll k_{B}T}\frac{\Mcal_z}{B_z}=\frac{2N(g\mu_B)^2}{k_B T}\left(\frac{1}{3+e^{-\zeta}}\right).
\label{bleaney}
\end{align}

From Eq. (\ref{Sup-ErgotropyCarboxylate}) is possible to write the ergotropy in the regime of magnetic susceptibility ($E_0\ll k_{B}T$ ) as
\begin{equation}
\Ecal_{E_0\ll k_{B}T}= \left\{
\begin{aligned}
E_{0} \left(\frac{1 - e^{{\zeta}}}{3e^{{\zeta}}+1}\right), ~~ E_{0}< {J} \\
0, ~~ E_{0}\geq {J}
\end{aligned}
\right.~.
\label{Ergotropysmall}
\end{equation}
As expected, there is no ergotropy for $E_{0}\geq {J}$, since in this region $B_{z}\geq B_{c}$ and the system cannot be found under the magnetic susceptibility regime. Thus, from Eqs. (\ref{bleaney}) and (\ref{Ergotropysmall}) it is possible to write the ergotropy in terms of the magnetic susceptibility of the compound as
\begin{equation}
\Ecal_{E_0\ll k_{B}T}(T)  = \
E_{0} \frac{k_BT\chi(T)}{2Ng^2\mu_B^2}\left[ e^{-\zeta} -1\right]~.
\label{Sup-Ergotropysus}
\end{equation}

Therefore, we establish a relationship between the available work stored in the quantum states of the reported metal complex and one of its thermodynamic properties. This results enables the measurement of the ergotropy of the system without disturbing the stored energy by any destructive readout.

\section{Quantumness of the battery}

One of the main tasks of this work is to explore the quantumness of our carboxylate based quantum battery, which is characterized by the set of physical quantities given by \textit{entanglement of formation} and \textit{quantum discord}:

\subsubsection{Entanglement of Formation}

Entanglement of formation is an entanglement quantifier which corresponds to the amount of quantum pure states needed in order to create a mixed entangled state. In this sense, entanglement in a statistical mixture would be the average of the mixture of pure states \cite{Hill:97,Wootters:98}. The entanglement of formation for a two-qubit system is defined as \cite{cruz,cruz2017influence}
\begin{equation}
\Fcal_{E}=-\mathbb{E}_+-\mathbb{E}_- ~~,
\label{entanglement}
\end{equation}
where 
\begin{equation}
\mathbb{E}_\pm=\frac{1\pm\sqrt{1-\mathbbm{C}^2}}{2}\log_2 \left(\frac{1\pm\sqrt{1-\mathbbm{C}^2}}{2}\right)
\end{equation}
and $\mathbbm{C}$ is the \textit{concurrence}~\cite{Hill:97,Wootters:98}, which is given by $\mathbbm{C}(\tau) = \max \{ 0, \lambda_{1} - \lambda_{2} - \lambda_{3} - \lambda_{4}\}$, where $\lambda_{1},\cdots,\lambda_{4}$ are the eigenvalues in decreasing order of the matrix $\hat R\!=\! ( \hat\rho^{1/2}\hat{\tilde{\rho}}\hat\rho^{1/2} )^{1/2}$, where $\hat{\tilde{\rho}}\!=\!(\hat\sigma_{y}\otimes \hat\sigma_{y}) \hat\rho^{*} (\hat\sigma_{y}\otimes \hat\sigma_{y})$, with $\hat\rho^{*}$ being the complex conjugate of $\hat\rho$ taken in the standard basis, which for our case is the battery energy basis. Thus, for our carboxylate based metal complex, \textit{concurrence} can be analytically computed as
\begin{align}
\mathbbm{C} = \max \left[0 ,\frac{e^{E_{0}\beta} \left( 1 - 3 e^{{\zeta}} \right)}{4\left( e^{{\zeta}} + e^{E_{0}\beta} + e^{{\zeta}+E_{0}\beta} + e^{{\zeta}+2E_{0}\beta} \right)}\right]~.
\label{concurence}
\end{align} % systems 

\subsubsection{Quantum discord}

Although entanglement provides a way to determine pure quantum correlations, it does not encompass all possible types of quantum correlations of a system \cite{cruz,Ciccarello:14,Obando:15,yurishchev2011quantum}. In this sense, quantum discord turns out to be a powerful tool for quantifying the quantumness of correlations \cite{Zurek:01}. We calculated the quantum discord from the trace distance discord, based on Schatten 1-norm ($||X||_{1}\!=\!\tr[\sqrt{X^{\dagger}X}]$), defined as the minimum distance between the state $\rho$ of a bipartite system and the set ($\Sigma$) of closest classical-quantum state $\sigma$, given by $\Dcal(\rho)\!=\!\min_{\sigma \in \Sigma}||\rho - \sigma ||_{1}$ \cite{Ciccarello:14}. For the system of interest, Eq.~\eqref{Sup-longitudinal}, the Schatten 1-norm quantum discord is given by
\begin{align}
\Dcal = \frac{1}{2} \left\vert\coth (\frac{\zeta}{2}) + \cosh (E_{0}\beta) \left( 1 + \coth (\frac{\zeta}{2}) \right)\right\vert^{-1}.
\label{Sup-discord}
\end{align}
%where one can remark that $\Dcal = 3C_{\ell_{1}}/2$, with $C_{\ell_{1}}$ being the quantum coherence defined from $\ell_{1}$ trace norm. Therefore, there is a direct connection between the quantum coherence, described by the non-diagonal terms of the density matrix, and the genuine quantum correlations between both spins of the compound.

\subsubsection{Ergotropy and Quantum Correlations}

\begin{figure}[t!]
	\includegraphics[scale=0.118]{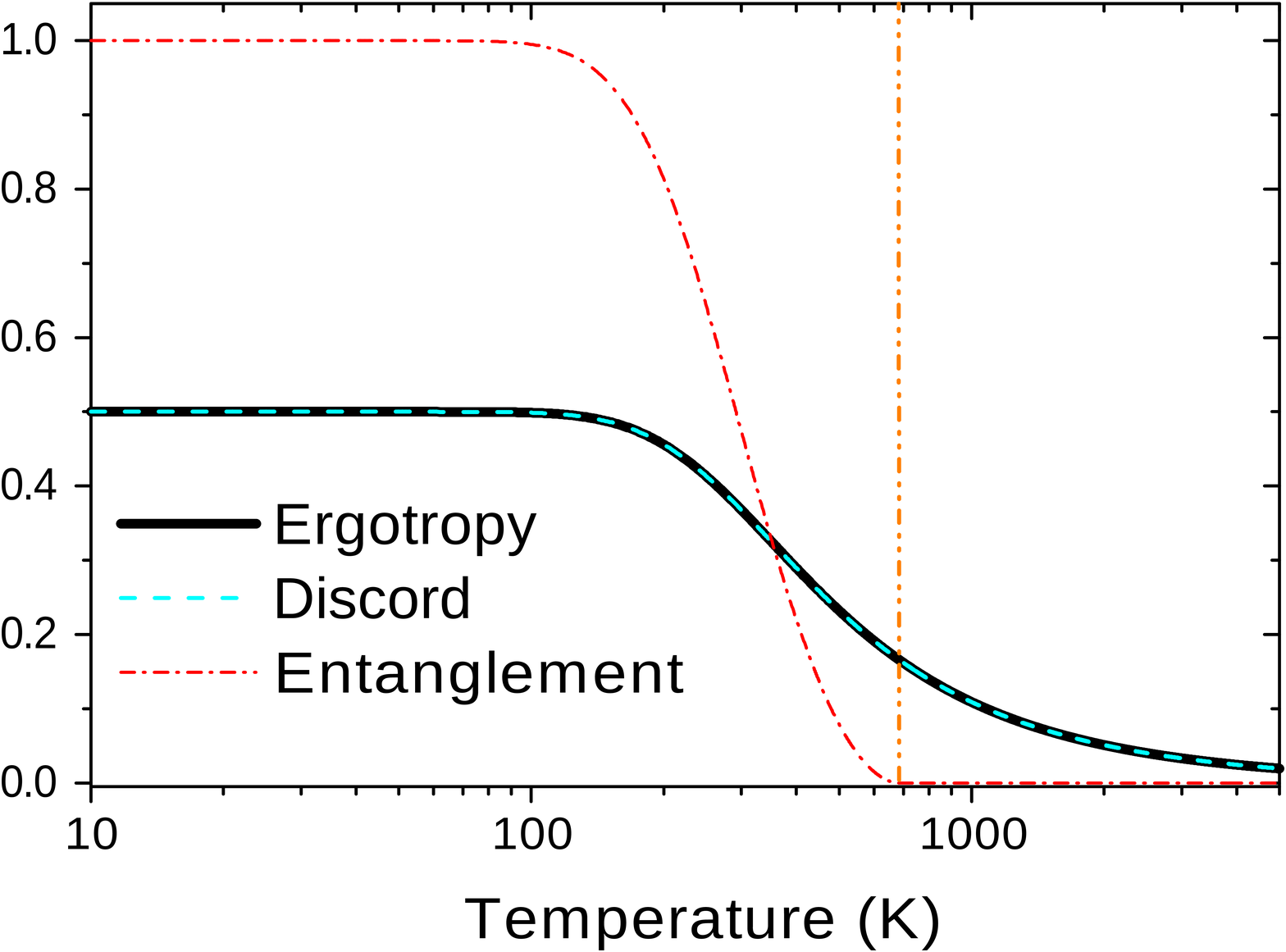}~
	\caption{(color online) (a) Entanglement of Formation (dot-dashed red line), Schatten 1-norm Quantum Discord (dashed blue line) and theoretical  dimensionless Ergotropy $\Ecal/\Ecal^{\text{max}}$(solid black line), as a function of the temperature for Earth's magnetic field $B_z\!=\!10^{-4}$ T. Vertical (orange) dotted line represents the temperature of entanglement $T_{\text{e}}=J /(\ln (3)k_{B})$ above which there is no entanglement ($\Fcal_{E}(T\!\geq\! T_{\text{e}})\!=\!0$).}
	\label{Sup-QuantitiesFig}
\end{figure}

Fig.~\ref{Sup-QuantitiesFig} shows all the quantities above described: entanglement of formation, quantum discord, and the dimensionless ergotropy  ($\Ecal/\Ecal^{\text{max}}$) as a function of temperature. The theoretical curves are plotted by Eqs.~\eqref{entanglement},~\eqref{Sup-discord} and~\eqref{Sup-ErgotropyCarboxylate}, respectively, considering the Earth's magnetic field of $B_z\!=\!10^{-4}$~T, and the experimental parameters $g=2.$ ($d^{9}$ ions) and $J/k_{B}\!=\!- 748.5$ K, in order to represent the laboratory environment in which batteries should work. As can be seen, the asymptotic decay of ergotropy with temperature coincides with the decay of Schatten 1-norm quantum discord, showing a strong relationship between these quantities, beyond entanglement. Hence, by the theoretically increasing of temperature, one finds that above the maximum  temperature $T_{\text{e}}=J /(\ln (3)k_{B})$, under which there is no entanglement ($\Fcal_{E}(T\!\geq\! T_{\text{e}})\!=\!0$), the ergotropy of the system is $16.6\%$ of the maximum one. This result suggests that quantum correlations beyond entanglement needs to taken into account to describe the amount of available work stored in metal complexes.

As pointed before, we can address the available work of our carboxylate-based quantum battery in the experimentally feasible region $E_{0}\!<\! J_{int}$  in the magnetic susceptibility regime, Eq. (\ref{Ergotropysus}). Applying the limit $E_0\ll k_{B}T$ in Eq. (\ref{Sup-discord}) we obtain
\begin{align}
\Dcal = \frac{1}{2} \left\vert\frac{1-e^{\zeta}}{3e^{\zeta} +1}\right\vert ~.
\label{discosus1}
\end{align}

Thus, comparing Eq. (\ref{Ergotropysmall}) with Eq. (\ref{discosus1}), we obtain 
\begin{equation} 
\Ecal_{E_0\ll k_{B}T}  = 2E_0 \Dcal ~~ \forall ~~ J>0 ~~,
\label{discord2}
\end{equation}
where $2E_0$ is the maximum energy stored, obtained from the spectral decomposition of the reference Hamiltonian $H_0$. 

Therefore, since ${\zeta}\!=\! -\beta J_{\text{int}}$, for any spin-$1/2$ quantum antiferromagnet ($J>0$), in the regime of magnetic susceptibility ($E_0\ll k_{B}T$), the existence of a finite amount of available work will be directly related to the presence of genuine quantum correlations between the cells of the battery. Then, the energy is stored from physical quantities and correlations that cannot be modeled by a classical system, i.e., we have a genuine QB.


\begin{thebibliography}{50}%
	\makeatletter
	\providecommand \@ifxundefined [1]{%
		\@ifx{#1\undefined}
	}%
	\providecommand \@ifnum [1]{%
		\ifnum #1\expandafter \@firstoftwo
		\else \expandafter \@secondoftwo
		\fi
	}%
	\providecommand \@ifx [1]{%
		\ifx #1\expandafter \@firstoftwo
		\else \expandafter \@secondoftwo
		\fi
	}%
	\providecommand \natexlab [1]{#1}%
	\providecommand \enquote  [1]{``#1''}%
	\providecommand \bibnamefont  [1]{#1}%
	\providecommand \bibfnamefont [1]{#1}%
	\providecommand \citenamefont [1]{#1}%
	\providecommand \href@noop [0]{\@secondoftwo}%
	\providecommand \href [0]{\begingroup \@sanitize@url \@href}%
	\providecommand \@href[1]{\@@startlink{#1}\@@href}%
	\providecommand \@@href[1]{\endgroup#1\@@endlink}%
	\providecommand \@sanitize@url [0]{\catcode `\\12\catcode `\$12\catcode
		`\&12\catcode `\#12\catcode `\^12\catcode `\_12\catcode `\%12\relax}%
	\providecommand \@@startlink[1]{}%
	\providecommand \@@endlink[0]{}%
	\providecommand \url  [0]{\begingroup\@sanitize@url \@url }%
	\providecommand \@url [1]{\endgroup\@href {#1}{\urlprefix }}%
	\providecommand \urlprefix  [0]{URL }%
	\providecommand \Eprint [0]{\href }%
	\providecommand \doibase [0]{http://dx.doi.org/}%
	\providecommand \selectlanguage [0]{\@gobble}%
	\providecommand \bibinfo  [0]{\@secondoftwo}%
	\providecommand \bibfield  [0]{\@secondoftwo}%
	\providecommand \translation [1]{[#1]}%
	\providecommand \BibitemOpen [0]{}%
	\providecommand \bibitemStop [0]{}%
	\providecommand \bibitemNoStop [0]{.\EOS\space}%
	\providecommand \EOS [0]{\spacefactor3000\relax}%
	\providecommand \BibitemShut  [1]{\csname bibitem#1\endcsname}%
	\let\auto@bib@innerbib\@empty
	%</preamble>
	\bibitem [{\citenamefont {Li}\ \emph {et~al.}(2018)\citenamefont {Li},
		\citenamefont {Lu}, \citenamefont {Chen},\ and\ \citenamefont
		{Amine}}]{li201830}%
	\BibitemOpen
	\bibfield  {author} {\bibinfo {author} {\bibfnamefont {M.}~\bibnamefont
			{Li}}, \bibinfo {author} {\bibfnamefont {J.}~\bibnamefont {Lu}}, \bibinfo
		{author} {\bibfnamefont {Z.}~\bibnamefont {Chen}}, \ and\ \bibinfo {author}
		{\bibfnamefont {K.}~\bibnamefont {Amine}},\ }\href {\doibase
		10.1002/adma.201800561} {\bibfield  {journal} {\bibinfo  {journal} {Advanced
				Materials}\ }\textbf {\bibinfo {volume} {30}},\ \bibinfo {pages} {1800561}
		(\bibinfo {year} {2018})}\BibitemShut {NoStop}%
	\bibitem [{\citenamefont {Liu}\ \emph {et~al.}(2019)\citenamefont {Liu},
		\citenamefont {Segal},\ and\ \citenamefont {Hanna}}]{Liu:19}%
	\BibitemOpen
	\bibfield  {author} {\bibinfo {author} {\bibfnamefont {J.}~\bibnamefont
			{Liu}}, \bibinfo {author} {\bibfnamefont {D.}~\bibnamefont {Segal}}, \ and\
		\bibinfo {author} {\bibfnamefont {G.}~\bibnamefont {Hanna}},\ }\href
	{\doibase 10.1021/acs.jpcc.9b06373} {\bibfield  {journal} {\bibinfo
			{journal} {J. Phys. Chem. C}\ }\textbf {\bibinfo {volume} {123}},\ \bibinfo
		{pages} {18303 } (\bibinfo {year} {2019})}\BibitemShut {NoStop}%
	\bibitem [{\citenamefont {Manzano-Agugliaro}\ \emph {et~al.}(2013)\citenamefont
		{Manzano-Agugliaro}, \citenamefont {Alcayde}, \citenamefont {Montoya},
		\citenamefont {Zapata-Sierra},\ and\ \citenamefont
		{Gil}}]{manzano2013scientific}%
	\BibitemOpen
	\bibfield  {author} {\bibinfo {author} {\bibfnamefont {F.}~\bibnamefont
			{Manzano-Agugliaro}}, \bibinfo {author} {\bibfnamefont {A.}~\bibnamefont
			{Alcayde}}, \bibinfo {author} {\bibfnamefont {F.~G.}\ \bibnamefont
			{Montoya}}, \bibinfo {author} {\bibfnamefont {A.}~\bibnamefont
			{Zapata-Sierra}}, \ and\ \bibinfo {author} {\bibfnamefont {C.}~\bibnamefont
			{Gil}},\ }\href {\doibase 10.1016/j.rser.2012.10.020} {\bibfield  {journal}
		{\bibinfo  {journal} {Renewable and Sustainable Energy Reviews}\ }\textbf
		{\bibinfo {volume} {18}},\ \bibinfo {pages} {134} (\bibinfo {year}
		{2013})}\BibitemShut {NoStop}%
	\bibitem [{\citenamefont {Ferraro}\ \emph {et~al.}(2018)\citenamefont
		{Ferraro}, \citenamefont {Campisi}, \citenamefont {Andolina}, \citenamefont
		{Pellegrini},\ and\ \citenamefont {Polini}}]{Ferraro:18}%
	\BibitemOpen
	\bibfield  {author} {\bibinfo {author} {\bibfnamefont {D.}~\bibnamefont
			{Ferraro}}, \bibinfo {author} {\bibfnamefont {M.}~\bibnamefont {Campisi}},
		\bibinfo {author} {\bibfnamefont {G.~M.}\ \bibnamefont {Andolina}}, \bibinfo
		{author} {\bibfnamefont {V.}~\bibnamefont {Pellegrini}}, \ and\ \bibinfo
		{author} {\bibfnamefont {M.}~\bibnamefont {Polini}},\ }\href {\doibase
		10.1103/PhysRevLett.120.117702} {\bibfield  {journal} {\bibinfo  {journal}
			{Phys. Rev. Lett.}\ }\textbf {\bibinfo {volume} {120}},\ \bibinfo {pages}
		{117702} (\bibinfo {year} {2018})}\BibitemShut {NoStop}%
	\bibitem [{\citenamefont {Alicki}\ and\ \citenamefont
		{Fannes}(2013)}]{Alicki:13}%
	\BibitemOpen
	\bibfield  {author} {\bibinfo {author} {\bibfnamefont {R.}~\bibnamefont
			{Alicki}}\ and\ \bibinfo {author} {\bibfnamefont {M.}~\bibnamefont
			{Fannes}},\ }\href {\doibase 10.1103/PhysRevE.87.042123} {\bibfield
		{journal} {\bibinfo  {journal} {Phys. Rev. E}\ }\textbf {\bibinfo {volume}
			{87}},\ \bibinfo {pages} {042123} (\bibinfo {year} {2013})}\BibitemShut
	{NoStop}%
	\bibitem [{\citenamefont {Binder}\ \emph {et~al.}(2015)\citenamefont {Binder},
		\citenamefont {Vinjanampathy}, \citenamefont {Modi},\ and\ \citenamefont
		{Goold}}]{Binder:15}%
	\BibitemOpen
	\bibfield  {author} {\bibinfo {author} {\bibfnamefont {F.~C.}\ \bibnamefont
			{Binder}}, \bibinfo {author} {\bibfnamefont {S.}~\bibnamefont
			{Vinjanampathy}}, \bibinfo {author} {\bibfnamefont {K.}~\bibnamefont {Modi}},
		\ and\ \bibinfo {author} {\bibfnamefont {J.}~\bibnamefont {Goold}},\ }\href
	{\doibase 10.1088/1367-2630/17/7/075015} {\bibfield  {journal} {\bibinfo
			{journal} {New J. Phys.}\ }\textbf {\bibinfo {volume} {17}},\ \bibinfo
		{pages} {075015} (\bibinfo {year} {2015})}\BibitemShut {NoStop}%
	\bibitem [{\citenamefont {Santos}\ \emph {et~al.}(2020)\citenamefont {Santos},
		\citenamefont {Saguia},\ and\ \citenamefont {Sarandy}}]{Santos:20c}%
	\BibitemOpen
	\bibfield  {author} {\bibinfo {author} {\bibfnamefont {A.~C.}\ \bibnamefont
			{Santos}}, \bibinfo {author} {\bibfnamefont {A.}~\bibnamefont {Saguia}}, \
		and\ \bibinfo {author} {\bibfnamefont {M.~S.}\ \bibnamefont {Sarandy}},\
	}\href {\doibase 10.1103/PhysRevE.101.062114} {\bibfield  {journal} {\bibinfo
			{journal} {Phys. Rev. E}\ }\textbf {\bibinfo {volume} {101}},\ \bibinfo
		{pages} {062114} (\bibinfo {year} {2020})}\BibitemShut {NoStop}%
	\bibitem [{\citenamefont {Giorgi}\ and\ \citenamefont
		{Campbell}(2015)}]{CampbellBatteries}%
	\BibitemOpen
	\bibfield  {author} {\bibinfo {author} {\bibfnamefont {G.~L.}\ \bibnamefont
			{Giorgi}}\ and\ \bibinfo {author} {\bibfnamefont {S.}~\bibnamefont
			{Campbell}},\ }\href {\doibase 10.1088/0953-4075/48/3/035501} {\bibfield
		{journal} {\bibinfo  {journal} {J. Phys. B: At. Mol. Opt. Phys.}\ }\textbf
		{\bibinfo {volume} {48}},\ \bibinfo {pages} {035501} (\bibinfo {year}
		{2015})}\BibitemShut {NoStop}%
	\bibitem [{\citenamefont {Campaioli}\ \emph {et~al.}(2017)\citenamefont
		{Campaioli}, \citenamefont {Pollock}, \citenamefont {Binder}, \citenamefont
		{C\'eleri}, \citenamefont {Goold}, \citenamefont {Vinjanampathy},\ and\
		\citenamefont {Modi}}]{PRL2017Binder}%
	\BibitemOpen
	\bibfield  {author} {\bibinfo {author} {\bibfnamefont {F.}~\bibnamefont
			{Campaioli}}, \bibinfo {author} {\bibfnamefont {F.~A.}\ \bibnamefont
			{Pollock}}, \bibinfo {author} {\bibfnamefont {F.~C.}\ \bibnamefont {Binder}},
		\bibinfo {author} {\bibfnamefont {L.}~\bibnamefont {C\'eleri}}, \bibinfo
		{author} {\bibfnamefont {J.}~\bibnamefont {Goold}}, \bibinfo {author}
		{\bibfnamefont {S.}~\bibnamefont {Vinjanampathy}}, \ and\ \bibinfo {author}
		{\bibfnamefont {K.}~\bibnamefont {Modi}},\ }\href {\doibase
		10.1103/PhysRevLett.118.150601} {\bibfield  {journal} {\bibinfo  {journal}
			{Phys. Rev. Lett.}\ }\textbf {\bibinfo {volume} {118}},\ \bibinfo {pages}
		{150601} (\bibinfo {year} {2017})}\BibitemShut {NoStop}%
	\bibitem [{\citenamefont {Andolina}\ \emph {et~al.}(2019)\citenamefont
		{Andolina}, \citenamefont {Keck}, \citenamefont {Mari}, \citenamefont
		{Campisi}, \citenamefont {Giovannetti},\ and\ \citenamefont
		{Polini}}]{PRL_Andolina}%
	\BibitemOpen
	\bibfield  {author} {\bibinfo {author} {\bibfnamefont {G.~M.}\ \bibnamefont
			{Andolina}}, \bibinfo {author} {\bibfnamefont {M.}~\bibnamefont {Keck}},
		\bibinfo {author} {\bibfnamefont {A.}~\bibnamefont {Mari}}, \bibinfo {author}
		{\bibfnamefont {M.}~\bibnamefont {Campisi}}, \bibinfo {author} {\bibfnamefont
			{V.}~\bibnamefont {Giovannetti}}, \ and\ \bibinfo {author} {\bibfnamefont
			{M.}~\bibnamefont {Polini}},\ }\href {\doibase
		10.1103/PhysRevLett.122.047702} {\bibfield  {journal} {\bibinfo  {journal}
			{Phys. Rev. Lett.}\ }\textbf {\bibinfo {volume} {122}},\ \bibinfo {pages}
		{047702} (\bibinfo {year} {2019})}\BibitemShut {NoStop}%
	\bibitem [{\citenamefont {{{\c{C}}akmak}}(2020)}]{Baris:20}%
	\BibitemOpen
	\bibfield  {author} {\bibinfo {author} {\bibfnamefont {B.}~\bibnamefont
			{{{\c{C}}akmak}}},\ }\href {\doibase 10.1103/PhysRevE.102.042111} {\bibfield
		{journal} {\bibinfo  {journal} {Phys. Rev. E}\ }\textbf {\bibinfo {volume}
			{102}},\ \bibinfo {pages} {042111} (\bibinfo {year} {2020})}\BibitemShut
	{NoStop}%
	\bibitem [{\citenamefont {Santos}\ \emph {et~al.}(2019)\citenamefont {Santos},
		\citenamefont {\c{C}akmak}, \citenamefont {Campbell},\ and\ \citenamefont
		{Zinner}}]{Santos:19-a}%
	\BibitemOpen
	\bibfield  {author} {\bibinfo {author} {\bibfnamefont {A.~C.}\ \bibnamefont
			{Santos}}, \bibinfo {author} {\bibfnamefont {B.}~\bibnamefont {\c{C}akmak}},
		\bibinfo {author} {\bibfnamefont {S.}~\bibnamefont {Campbell}}, \ and\
		\bibinfo {author} {\bibfnamefont {N.~T.}\ \bibnamefont {Zinner}},\ }\href
	{\doibase 10.1103/PhysRevE.100.032107} {\bibfield  {journal} {\bibinfo
			{journal} {Phys. Rev. E}\ }\textbf {\bibinfo {volume} {100}},\ \bibinfo
		{pages} {032107} (\bibinfo {year} {2019})}\BibitemShut {NoStop}%
	\bibitem [{\citenamefont {{\c{C}}akmak}(2020)}]{PhysRevE.102.042111}%
	\BibitemOpen
	\bibfield  {author} {\bibinfo {author} {\bibfnamefont {B.}~\bibnamefont
			{{\c{C}}akmak}},\ }\href {\doibase 10.1103/PhysRevE.102.042111} {\bibfield
		{journal} {\bibinfo  {journal} {Phys. Rev. E}\ }\textbf {\bibinfo {volume}
			{102}},\ \bibinfo {pages} {042111} (\bibinfo {year} {2020})}\BibitemShut
	{NoStop}%
	\bibitem [{\citenamefont {Francica}\ \emph {et~al.}(2020)\citenamefont
		{Francica}, \citenamefont {Binder}, \citenamefont {Guarnieri}, \citenamefont
		{Mitchison}, \citenamefont {Goold},\ and\ \citenamefont
		{Plastina}}]{PhysRevLett.125.180603}%
	\BibitemOpen
	\bibfield  {author} {\bibinfo {author} {\bibfnamefont {G.}~\bibnamefont
			{Francica}}, \bibinfo {author} {\bibfnamefont {F.~C.}\ \bibnamefont
			{Binder}}, \bibinfo {author} {\bibfnamefont {G.}~\bibnamefont {Guarnieri}},
		\bibinfo {author} {\bibfnamefont {M.~T.}\ \bibnamefont {Mitchison}}, \bibinfo
		{author} {\bibfnamefont {J.}~\bibnamefont {Goold}}, \ and\ \bibinfo {author}
		{\bibfnamefont {F.}~\bibnamefont {Plastina}},\ }\href {\doibase
		10.1103/PhysRevLett.125.180603} {\bibfield  {journal} {\bibinfo  {journal}
			{Phys. Rev. Lett.}\ }\textbf {\bibinfo {volume} {125}},\ \bibinfo {pages}
		{180603} (\bibinfo {year} {2020})}\BibitemShut {NoStop}%
	\bibitem [{\citenamefont {Le}\ \emph {et~al.}(2018)\citenamefont {Le},
		\citenamefont {Levinsen}, \citenamefont {Modi}, \citenamefont {Parish},\ and\
		\citenamefont {Pollock}}]{Le:18}%
	\BibitemOpen
	\bibfield  {author} {\bibinfo {author} {\bibfnamefont {T.~P.}\ \bibnamefont
			{Le}}, \bibinfo {author} {\bibfnamefont {J.}~\bibnamefont {Levinsen}},
		\bibinfo {author} {\bibfnamefont {K.}~\bibnamefont {Modi}}, \bibinfo {author}
		{\bibfnamefont {M.~M.}\ \bibnamefont {Parish}}, \ and\ \bibinfo {author}
		{\bibfnamefont {F.~A.}\ \bibnamefont {Pollock}},\ }\href {\doibase
		10.1103/PhysRevA.97.022106} {\bibfield  {journal} {\bibinfo  {journal} {Phys.
				Rev. A}\ }\textbf {\bibinfo {volume} {97}},\ \bibinfo {pages} {022106}
		(\bibinfo {year} {2018})}\BibitemShut {NoStop}%
	\bibitem [{\citenamefont {Strambini}\ \emph {et~al.}(2020)\citenamefont
		{Strambini}, \citenamefont {Iorio}, \citenamefont {Durante}, \citenamefont
		{Citro}, \citenamefont {Sanz-Fern{\'a}ndez}, \citenamefont {Guarcello},
		\citenamefont {Tokatly}, \citenamefont {Braggio}, \citenamefont {Rocci},
		\citenamefont {Ligato} \emph {et~al.}}]{strambini2020josephson}%
	\BibitemOpen
	\bibfield  {author} {\bibinfo {author} {\bibfnamefont {E.}~\bibnamefont
			{Strambini}}, \bibinfo {author} {\bibfnamefont {A.}~\bibnamefont {Iorio}},
		\bibinfo {author} {\bibfnamefont {O.}~\bibnamefont {Durante}}, \bibinfo
		{author} {\bibfnamefont {R.}~\bibnamefont {Citro}}, \bibinfo {author}
		{\bibfnamefont {C.}~\bibnamefont {Sanz-Fern{\'a}ndez}}, \bibinfo {author}
		{\bibfnamefont {C.}~\bibnamefont {Guarcello}}, \bibinfo {author}
		{\bibfnamefont {I.~V.}\ \bibnamefont {Tokatly}}, \bibinfo {author}
		{\bibfnamefont {A.}~\bibnamefont {Braggio}}, \bibinfo {author} {\bibfnamefont
			{M.}~\bibnamefont {Rocci}}, \bibinfo {author} {\bibfnamefont
			{N.}~\bibnamefont {Ligato}},  \emph {et~al.},\ }\href {\doibase
		10.1038/s41565-020-0712-7} {\bibfield  {journal} {\bibinfo  {journal} {Nature
				Nanotechnology}\ }\textbf {\bibinfo {volume} {15}},\ \bibinfo {pages} {656}
		(\bibinfo {year} {2020})}\BibitemShut {NoStop}%
	\bibitem [{\citenamefont {Rossini}\ \emph {et~al.}(2020)\citenamefont
		{Rossini}, \citenamefont {Andolina}, \citenamefont {Rosa}, \citenamefont
		{Carrega},\ and\ \citenamefont {Polini}}]{Rossini:20}%
	\BibitemOpen
	\bibfield  {author} {\bibinfo {author} {\bibfnamefont {D.}~\bibnamefont
			{Rossini}}, \bibinfo {author} {\bibfnamefont {G.~M.}\ \bibnamefont
			{Andolina}}, \bibinfo {author} {\bibfnamefont {D.}~\bibnamefont {Rosa}},
		\bibinfo {author} {\bibfnamefont {M.}~\bibnamefont {Carrega}}, \ and\
		\bibinfo {author} {\bibfnamefont {M.}~\bibnamefont {Polini}},\ }\href
	{\doibase 10.1103/PhysRevLett.125.236402} {\bibfield  {journal} {\bibinfo
			{journal} {Phys. Rev. Lett.}\ }\textbf {\bibinfo {volume} {125}},\ \bibinfo
		{pages} {236402} (\bibinfo {year} {2020})}\BibitemShut {NoStop}%
	\bibitem [{\citenamefont {Monsel}\ \emph {et~al.}(2020)\citenamefont {Monsel},
		\citenamefont {Fellous-Asiani}, \citenamefont {Huard},\ and\ \citenamefont
		{Auff\`eves}}]{Alexia:20}%
	\BibitemOpen
	\bibfield  {author} {\bibinfo {author} {\bibfnamefont {J.}~\bibnamefont
			{Monsel}}, \bibinfo {author} {\bibfnamefont {M.}~\bibnamefont
			{Fellous-Asiani}}, \bibinfo {author} {\bibfnamefont {B.}~\bibnamefont
			{Huard}}, \ and\ \bibinfo {author} {\bibfnamefont {A.}~\bibnamefont
			{Auff\`eves}},\ }\href {\doibase 10.1103/PhysRevLett.124.130601} {\bibfield
		{journal} {\bibinfo  {journal} {Phys. Rev. Lett.}\ }\textbf {\bibinfo
			{volume} {124}},\ \bibinfo {pages} {130601} (\bibinfo {year}
		{2020})}\BibitemShut {NoStop}%
	\bibitem [{ccd()}]{ccdc}%
	\BibitemOpen
	\href@noop {} {}\bibinfo {note} {CCDC 1428140 contains the supplementary
		crystallographic data for this paper. These data can be obtained free of
		charge from The Cambridge Crystallographic Data Centre via
		\url{www.ccdc.cam.ac.uk/structures}}\BibitemShut {NoStop}%
	\bibitem [{ves()}]{vesta}%
	\BibitemOpen
	\href@noop {} {}\bibinfo {note} {Visualization of the structures was made
		using the Visualisation for Electronic Structural Analysis (VESTA) software
		\cite{momma2011vesta}.}\BibitemShut {Stop}%
	\bibitem [{\citenamefont {Cruz}\ \emph {et~al.}(2016)\citenamefont {Cruz},
		\citenamefont {Soares-Pinto}, \citenamefont {{a}o}, \citenamefont {dos
			Santos},\ and\ \citenamefont {Reis}}]{cruz}%
	\BibitemOpen
	\bibfield  {author} {\bibinfo {author} {\bibfnamefont {C.}~\bibnamefont
			{Cruz}}, \bibinfo {author} {\bibfnamefont {D.~O.}\ \bibnamefont
			{Soares-Pinto}}, \bibinfo {author} {\bibfnamefont {P.~B.}\ \bibnamefont
			{{a}o}}, \bibinfo {author} {\bibfnamefont {A.~M.}\ \bibnamefont {dos
				Santos}}, \ and\ \bibinfo {author} {\bibfnamefont {M.~S.}\ \bibnamefont
			{Reis}},\ }\href {\doibase 10.1209/0295-5075/113/40004} {\bibfield  {journal}
		{\bibinfo  {journal} {EPL (Europhysics Letters)}\ }\textbf {\bibinfo {volume}
			{113}},\ \bibinfo {pages} {40004} (\bibinfo {year} {2016})}\BibitemShut
	{NoStop}%
	\bibitem [{\citenamefont {Souza}\ \emph {et~al.}(2009)\citenamefont {Souza},
		\citenamefont {Soares-Pinto}, \citenamefont {Sarthour}, \citenamefont
		{Oliveira}, \citenamefont {Reis}, \citenamefont {Brandao},\ and\
		\citenamefont {dos Santos}}]{souza}%
	\BibitemOpen
	\bibfield  {author} {\bibinfo {author} {\bibfnamefont {A.~M.}\ \bibnamefont
			{Souza}}, \bibinfo {author} {\bibfnamefont {D.~O.}\ \bibnamefont
			{Soares-Pinto}}, \bibinfo {author} {\bibfnamefont {R.~S.}\ \bibnamefont
			{Sarthour}}, \bibinfo {author} {\bibfnamefont {I.~S.}\ \bibnamefont
			{Oliveira}}, \bibinfo {author} {\bibfnamefont {M.~S.}\ \bibnamefont {Reis}},
		\bibinfo {author} {\bibfnamefont {P.}~\bibnamefont {Brandao}}, \ and\
		\bibinfo {author} {\bibfnamefont {A.~M.}\ \bibnamefont {dos Santos}},\ }\href
	{\doibase 10.1103/PhysRevB.79.054408} {\bibfield  {journal} {\bibinfo
			{journal} {Physical Review B}\ }\textbf {\bibinfo {volume} {79}},\ \bibinfo
		{pages} {054408} (\bibinfo {year} {2009})}\BibitemShut {NoStop}%
	\bibitem [{\citenamefont {Reis}\ \emph {et~al.}(2012)\citenamefont {Reis},
		\citenamefont {Soriano}, \citenamefont {dos Santos}, \citenamefont {Sales},
		\citenamefont {Soares-Pinto},\ and\ \citenamefont {Brandao}}]{mario2}%
	\BibitemOpen
	\bibfield  {author} {\bibinfo {author} {\bibfnamefont {M.~S.}\ \bibnamefont
			{Reis}}, \bibinfo {author} {\bibfnamefont {S.}~\bibnamefont {Soriano}},
		\bibinfo {author} {\bibfnamefont {A.~M.}\ \bibnamefont {dos Santos}},
		\bibinfo {author} {\bibfnamefont {B.~C.}\ \bibnamefont {Sales}}, \bibinfo
		{author} {\bibfnamefont {D.}~\bibnamefont {Soares-Pinto}}, \ and\ \bibinfo
		{author} {\bibfnamefont {P.}~\bibnamefont {Brandao}},\ }\href {\doibase
		10.1209/0295-5075/100/50001} {\bibfield  {journal} {\bibinfo  {journal} {EPL
				(Europhysics Letters)}\ }\textbf {\bibinfo {volume} {100}},\ \bibinfo {pages}
		{50001} (\bibinfo {year} {2012})}\BibitemShut {NoStop}%
	\bibitem [{\citenamefont {{\v{C}}en{\v{c}}arikov{\'a}}\ and\ \citenamefont
		{Stre{\v{c}}ka}(2020)}]{kova2020unconventional}%
	\BibitemOpen
	\bibfield  {author} {\bibinfo {author} {\bibfnamefont {H.}~\bibnamefont
			{{\v{C}}en{\v{c}}arikov{\'a}}}\ and\ \bibinfo {author} {\bibfnamefont
			{J.}~\bibnamefont {Stre{\v{c}}ka}},\ }\href {\doibase
		10.1103/PhysRevB.102.184419} {\bibfield  {journal} {\bibinfo  {journal}
			{Physical Review B}\ }\textbf {\bibinfo {volume} {102}},\ \bibinfo {pages}
		{184419} (\bibinfo {year} {2020})}\BibitemShut {NoStop}%
	\bibitem [{\citenamefont {Cruz}\ and\ \citenamefont
		{Anka}(2020)}]{cruz2020quantifying}%
	\BibitemOpen
	\bibfield  {author} {\bibinfo {author} {\bibfnamefont {C.}~\bibnamefont
			{Cruz}}\ and\ \bibinfo {author} {\bibfnamefont {M.}~\bibnamefont {Anka}},\
	}\href {\doibase 10.1209/0295-5075/130/30006} {\bibfield  {journal} {\bibinfo
			{journal} {EPL (Europhysics Letters)}\ }\textbf {\bibinfo {volume} {130}},\
		\bibinfo {pages} {30006} (\bibinfo {year} {2020})}\BibitemShut {NoStop}%
	\bibitem [{\citenamefont {Souza}\ \emph {et~al.}(2008)\citenamefont {Souza},
		\citenamefont {Reis}, \citenamefont {Soares-Pinto}, \citenamefont
		{Oliveira},\ and\ \citenamefont {Sarthour}}]{souza2}%
	\BibitemOpen
	\bibfield  {author} {\bibinfo {author} {\bibfnamefont {A.~M.}\ \bibnamefont
			{Souza}}, \bibinfo {author} {\bibfnamefont {M.~S.}\ \bibnamefont {Reis}},
		\bibinfo {author} {\bibfnamefont {D.~O.}\ \bibnamefont {Soares-Pinto}},
		\bibinfo {author} {\bibfnamefont {I.~S.}\ \bibnamefont {Oliveira}}, \ and\
		\bibinfo {author} {\bibfnamefont {R.~S.}\ \bibnamefont {Sarthour}},\ }\href
	{\doibase 10.1103/PhysRevB.77.104402} {\bibfield  {journal} {\bibinfo
			{journal} {Physical Review B}\ }\textbf {\bibinfo {volume} {77}},\ \bibinfo
		{pages} {104402} (\bibinfo {year} {2008})}\BibitemShut {NoStop}%
	\bibitem [{\citenamefont {He}\ \emph {et~al.}(2017)\citenamefont {He},
		\citenamefont {Jiang}, \citenamefont {Yu}, \citenamefont {Lin},\ and\
		\citenamefont {Guan}}]{he2017quantum}%
	\BibitemOpen
	\bibfield  {author} {\bibinfo {author} {\bibfnamefont {F.}~\bibnamefont
			{He}}, \bibinfo {author} {\bibfnamefont {Y.}~\bibnamefont {Jiang}}, \bibinfo
		{author} {\bibfnamefont {Y.-C.}\ \bibnamefont {Yu}}, \bibinfo {author}
		{\bibfnamefont {H.-Q.}\ \bibnamefont {Lin}}, \ and\ \bibinfo {author}
		{\bibfnamefont {X.-W.}\ \bibnamefont {Guan}},\ }\href {\doibase
		10.1103/PhysRevB.96.220401} {\bibfield  {journal} {\bibinfo  {journal}
			{Physical Review B}\ }\textbf {\bibinfo {volume} {96}},\ \bibinfo {pages}
		{220401} (\bibinfo {year} {2017})}\BibitemShut {NoStop}%
	\bibitem [{\citenamefont {Breunig}\ \emph {et~al.}(2017)\citenamefont
		{Breunig}, \citenamefont {Garst}, \citenamefont {Kl{\"u}mper}, \citenamefont
		{Rohrkamp}, \citenamefont {Turnbull},\ and\ \citenamefont
		{Lorenz}}]{breunig2017quantum}%
	\BibitemOpen
	\bibfield  {author} {\bibinfo {author} {\bibfnamefont {O.}~\bibnamefont
			{Breunig}}, \bibinfo {author} {\bibfnamefont {M.}~\bibnamefont {Garst}},
		\bibinfo {author} {\bibfnamefont {A.}~\bibnamefont {Kl{\"u}mper}}, \bibinfo
		{author} {\bibfnamefont {J.}~\bibnamefont {Rohrkamp}}, \bibinfo {author}
		{\bibfnamefont {M.~M.}\ \bibnamefont {Turnbull}}, \ and\ \bibinfo {author}
		{\bibfnamefont {T.}~\bibnamefont {Lorenz}},\ }\href {\doibase
		10.1126/sciadv.aao3773} {\bibfield  {journal} {\bibinfo  {journal} {Science
				advances}\ }\textbf {\bibinfo {volume} {3}},\ \bibinfo {pages} {eaao3773}
		(\bibinfo {year} {2017})}\BibitemShut {NoStop}%
	\bibitem [{\citenamefont {Cruz}\ \emph {et~al.}(2017)\citenamefont {Cruz},
		\citenamefont {Alves}, \citenamefont {dos Santos}, \citenamefont
		{Soares-Pinto}, \citenamefont {de~Jesus}, \citenamefont {de~Almeida},\ and\
		\citenamefont {Reis}}]{cruz2017influence}%
	\BibitemOpen
	\bibfield  {author} {\bibinfo {author} {\bibfnamefont {C.}~\bibnamefont
			{Cruz}}, \bibinfo {author} {\bibfnamefont {{\'A}.}~\bibnamefont {Alves}},
		\bibinfo {author} {\bibfnamefont {R.}~\bibnamefont {dos Santos}}, \bibinfo
		{author} {\bibfnamefont {D.}~\bibnamefont {Soares-Pinto}}, \bibinfo {author}
		{\bibfnamefont {J.}~\bibnamefont {de~Jesus}}, \bibinfo {author}
		{\bibfnamefont {J.}~\bibnamefont {de~Almeida}}, \ and\ \bibinfo {author}
		{\bibfnamefont {M.}~\bibnamefont {Reis}},\ }\href {\doibase
		10.1209/0295-5075/117/20004} {\bibfield  {journal} {\bibinfo  {journal} {EPL
				(Europhysics Letters)}\ }\textbf {\bibinfo {volume} {117}},\ \bibinfo {pages}
		{20004} (\bibinfo {year} {2017})}\BibitemShut {NoStop}%
	\bibitem [{\citenamefont {{Santos}}(2021)}]{Santos:20d}%
	\BibitemOpen
	\bibfield  {author} {\bibinfo {author} {\bibfnamefont {A.~C.}\ \bibnamefont
			{{Santos}}},\ }\href@noop {} {\bibfield  {journal} {\bibinfo  {journal}
			{Accepted for publication in Phys. Rev. E}\ } (\bibinfo {year} {2021})},\
	\Eprint {http://arxiv.org/abs/2012.11996} {arXiv:2012.11996 [quant-ph]}
	\BibitemShut {NoStop}%
	\bibitem [{\citenamefont {Yurishchev}(2011)}]{yurishchev2011quantum}%
	\BibitemOpen
	\bibfield  {author} {\bibinfo {author} {\bibfnamefont {M.~A.}\ \bibnamefont
			{Yurishchev}},\ }\href {\doibase 10.1103/PhysRevB.84.024418} {\bibfield
		{journal} {\bibinfo  {journal} {Physical Review B}\ }\textbf {\bibinfo
			{volume} {84}},\ \bibinfo {pages} {024418} (\bibinfo {year}
		{2011})}\BibitemShut {NoStop}%
	\bibitem [{\citenamefont {Gaita-Ari{\~n}o}\ \emph {et~al.}(2019)\citenamefont
		{Gaita-Ari{\~n}o}, \citenamefont {Luis}, \citenamefont {Hill},\ and\
		\citenamefont {Coronado}}]{gaita2019molecular}%
	\BibitemOpen
	\bibfield  {author} {\bibinfo {author} {\bibfnamefont {A.}~\bibnamefont
			{Gaita-Ari{\~n}o}}, \bibinfo {author} {\bibfnamefont {F.}~\bibnamefont
			{Luis}}, \bibinfo {author} {\bibfnamefont {S.}~\bibnamefont {Hill}}, \ and\
		\bibinfo {author} {\bibfnamefont {E.}~\bibnamefont {Coronado}},\ }\href@noop
	{} {\bibfield  {journal} {\bibinfo  {journal} {Nature chemistry}\ }\textbf
		{\bibinfo {volume} {11}},\ \bibinfo {pages} {301} (\bibinfo {year}
		{2019})}\BibitemShut {NoStop}%
	\bibitem [{\citenamefont {Moreno-Pineda}\ \emph {et~al.}(2018)\citenamefont
		{Moreno-Pineda}, \citenamefont {Godfrin}, \citenamefont {Balestro},
		\citenamefont {Wernsdorfer},\ and\ \citenamefont
		{Ruben}}]{moreno2018molecular}%
	\BibitemOpen
	\bibfield  {author} {\bibinfo {author} {\bibfnamefont {E.}~\bibnamefont
			{Moreno-Pineda}}, \bibinfo {author} {\bibfnamefont {C.}~\bibnamefont
			{Godfrin}}, \bibinfo {author} {\bibfnamefont {F.}~\bibnamefont {Balestro}},
		\bibinfo {author} {\bibfnamefont {W.}~\bibnamefont {Wernsdorfer}}, \ and\
		\bibinfo {author} {\bibfnamefont {M.}~\bibnamefont {Ruben}},\ }\href
	{\doibase 10.1039/C5CS00933B} {\bibfield  {journal} {\bibinfo  {journal}
			{Chemical Society Reviews}\ }\textbf {\bibinfo {volume} {47}},\ \bibinfo
		{pages} {501} (\bibinfo {year} {2018})}\BibitemShut {NoStop}%
	\bibitem [{\citenamefont {Chakraborty}\ and\ \citenamefont
		{Mitra}(2019)}]{chakraborty2019magnetocaloric}%
	\BibitemOpen
	\bibfield  {author} {\bibinfo {author} {\bibfnamefont {T.}~\bibnamefont
			{Chakraborty}}\ and\ \bibinfo {author} {\bibfnamefont {C.}~\bibnamefont
			{Mitra}},\ }\href {\doibase 10.1088/1361-648X/ab3962} {\bibfield  {journal}
		{\bibinfo  {journal} {Journal of Physics: Condensed Matter}\ }\textbf
		{\bibinfo {volume} {31}},\ \bibinfo {pages} {475802} (\bibinfo {year}
		{2019})}\BibitemShut {NoStop}%
	\bibitem [{\citenamefont {Reis}(2013)}]{mario}%
	\BibitemOpen
	\bibfield  {author} {\bibinfo {author} {\bibfnamefont {M.}~\bibnamefont
			{Reis}},\ }\href@noop {} {\emph {\bibinfo {title} {Fundamentals of
				magnetism}}}\ (\bibinfo  {publisher} {Elsevier},\ \bibinfo {year}
	{2013})\BibitemShut {NoStop}%
	\bibitem [{Sup()}]{SupInf}%
	\BibitemOpen
	\href@noop {} {}\bibinfo {note} {See Supplemental Material}\BibitemShut
	{NoStop}%
	\bibitem [{\citenamefont {Allahverdyan}\ \emph {et~al.}(2004)\citenamefont
		{Allahverdyan}, \citenamefont {Balian},\ and\ \citenamefont
		{Nieuwenhuizen}}]{Allahverdyan:04}%
	\BibitemOpen
	\bibfield  {author} {\bibinfo {author} {\bibfnamefont {A.~E.}\ \bibnamefont
			{Allahverdyan}}, \bibinfo {author} {\bibfnamefont {R.}~\bibnamefont
			{Balian}}, \ and\ \bibinfo {author} {\bibfnamefont {T.~M.}\ \bibnamefont
			{Nieuwenhuizen}},\ }\href {\doibase 10.1209/epl/i2004-10101-2} {\bibfield
		{journal} {\bibinfo  {journal} {Europhys. Lett.}\ }\textbf {\bibinfo {volume}
			{67}},\ \bibinfo {pages} {565} (\bibinfo {year} {2004})}\BibitemShut
	{NoStop}%
	\bibitem [{\citenamefont {Bleaney}\ and\ \citenamefont
		{Bowers}(1952)}]{bleaney1952anomalous}%
	\BibitemOpen
	\bibfield  {author} {\bibinfo {author} {\bibfnamefont {B.}~\bibnamefont
			{Bleaney}}\ and\ \bibinfo {author} {\bibfnamefont {K.}~\bibnamefont
			{Bowers}},\ }\href {\doibase 10.1098/rspa.1952.0181} {\bibfield  {journal}
		{\bibinfo  {journal} {Proceedings of the Royal Society of London. Series A.
				Mathematical and Physical Sciences}\ }\textbf {\bibinfo {volume} {214}},\
		\bibinfo {pages} {451} (\bibinfo {year} {1952})}\BibitemShut {NoStop}%
	\bibitem [{\citenamefont {Hill}\ and\ \citenamefont
		{Wootters}(1997)}]{Hill:97}%
	\BibitemOpen
	\bibfield  {author} {\bibinfo {author} {\bibfnamefont {S.}~\bibnamefont
			{Hill}}\ and\ \bibinfo {author} {\bibfnamefont {W.~K.}\ \bibnamefont
			{Wootters}},\ }\href {\doibase 10.1103/PhysRevLett.78.5022} {\bibfield
		{journal} {\bibinfo  {journal} {Phys. Rev. Lett.}\ }\textbf {\bibinfo
			{volume} {78}},\ \bibinfo {pages} {5022} (\bibinfo {year}
		{1997})}\BibitemShut {NoStop}%
	\bibitem [{\citenamefont {Wootters}(1998)}]{Wootters:98}%
	\BibitemOpen
	\bibfield  {author} {\bibinfo {author} {\bibfnamefont {W.~K.}\ \bibnamefont
			{Wootters}},\ }\href {\doibase 10.1103/PhysRevLett.80.2245} {\bibfield
		{journal} {\bibinfo  {journal} {Phys. Rev. Lett.}\ }\textbf {\bibinfo
			{volume} {80}},\ \bibinfo {pages} {2245} (\bibinfo {year}
		{1998})}\BibitemShut {NoStop}%
	\bibitem [{\citenamefont {Ciccarello}\ \emph {et~al.}(2014)\citenamefont
		{Ciccarello}, \citenamefont {Tufarelli},\ and\ \citenamefont
		{Giovannetti}}]{Ciccarello:14}%
	\BibitemOpen
	\bibfield  {author} {\bibinfo {author} {\bibfnamefont {F.}~\bibnamefont
			{Ciccarello}}, \bibinfo {author} {\bibfnamefont {T.}~\bibnamefont
			{Tufarelli}}, \ and\ \bibinfo {author} {\bibfnamefont {V.}~\bibnamefont
			{Giovannetti}},\ }\href {\doibase 10.1088/1367-2630/16/1/013038} {\bibfield
		{journal} {\bibinfo  {journal} {New J. Phys.}\ }\textbf {\bibinfo {volume}
			{16}},\ \bibinfo {pages} {013038} (\bibinfo {year} {2014})}\BibitemShut
	{NoStop}%
	\bibitem [{\citenamefont {Kwon}\ \emph {et~al.}(2018)\citenamefont {Kwon},
		\citenamefont {Jeong}, \citenamefont {Jennings}, \citenamefont {Yadin},\ and\
		\citenamefont {Kim}}]{Kwon:18}%
	\BibitemOpen
	\bibfield  {author} {\bibinfo {author} {\bibfnamefont {H.}~\bibnamefont
			{Kwon}}, \bibinfo {author} {\bibfnamefont {H.}~\bibnamefont {Jeong}},
		\bibinfo {author} {\bibfnamefont {D.}~\bibnamefont {Jennings}}, \bibinfo
		{author} {\bibfnamefont {B.}~\bibnamefont {Yadin}}, \ and\ \bibinfo {author}
		{\bibfnamefont {M.~S.}\ \bibnamefont {Kim}},\ }\href {\doibase
		10.1103/PhysRevLett.120.150602} {\bibfield  {journal} {\bibinfo  {journal}
			{Phys. Rev. Lett.}\ }\textbf {\bibinfo {volume} {120}},\ \bibinfo {pages}
		{150602} (\bibinfo {year} {2018})}\BibitemShut {NoStop}%
	\bibitem [{\citenamefont {Kamin}\ \emph {et~al.}(2020)\citenamefont {Kamin},
		\citenamefont {Tabesh}, \citenamefont {Salimi},\ and\ \citenamefont
		{Santos}}]{Kamin:20-2}%
	\BibitemOpen
	\bibfield  {author} {\bibinfo {author} {\bibfnamefont {F.~H.}\ \bibnamefont
			{Kamin}}, \bibinfo {author} {\bibfnamefont {F.~T.}\ \bibnamefont {Tabesh}},
		\bibinfo {author} {\bibfnamefont {S.}~\bibnamefont {Salimi}}, \ and\ \bibinfo
		{author} {\bibfnamefont {A.~C.}\ \bibnamefont {Santos}},\ }\href {\doibase
		10.1103/PhysRevE.102.052109} {\bibfield  {journal} {\bibinfo  {journal}
			{Phys. Rev. E}\ }\textbf {\bibinfo {volume} {102}},\ \bibinfo {pages}
		{052109} (\bibinfo {year} {2020})}\BibitemShut {NoStop}%
	\bibitem [{\citenamefont {Hovhannisyan}\ \emph {et~al.}(2020)\citenamefont
		{Hovhannisyan}, \citenamefont {Barra},\ and\ \citenamefont
		{Imparato}}]{Hovhannisyan:20}%
	\BibitemOpen
	\bibfield  {author} {\bibinfo {author} {\bibfnamefont {K.~V.}\ \bibnamefont
			{Hovhannisyan}}, \bibinfo {author} {\bibfnamefont {F.}~\bibnamefont {Barra}},
		\ and\ \bibinfo {author} {\bibfnamefont {A.}~\bibnamefont {Imparato}},\
	}\href {\doibase 10.1103/PhysRevResearch.2.033413} {\bibfield  {journal}
		{\bibinfo  {journal} {Phys. Rev. Research}\ }\textbf {\bibinfo {volume}
			{2}},\ \bibinfo {pages} {033413} (\bibinfo {year} {2020})}\BibitemShut
	{NoStop}%
	\bibitem [{\citenamefont {Sapienza}\ \emph {et~al.}(2019)\citenamefont
		{Sapienza}, \citenamefont {Cerisola},\ and\ \citenamefont
		{Roncaglia}}]{Sapienza:19}%
	\BibitemOpen
	\bibfield  {author} {\bibinfo {author} {\bibfnamefont {F.}~\bibnamefont
			{Sapienza}}, \bibinfo {author} {\bibfnamefont {F.}~\bibnamefont {Cerisola}},
		\ and\ \bibinfo {author} {\bibfnamefont {A.~J.}\ \bibnamefont {Roncaglia}},\
	}\href {\doibase 10.1038/s41467-019-10572-8} {\bibfield  {journal} {\bibinfo
			{journal} {Nature communications}\ }\textbf {\bibinfo {volume} {10}},\
		\bibinfo {pages} {1} (\bibinfo {year} {2019})}\BibitemShut {NoStop}%
	\bibitem [{\citenamefont {Wasielewski}\ \emph {et~al.}(2020)\citenamefont
		{Wasielewski}, \citenamefont {Forbes}, \citenamefont {Frank}, \citenamefont
		{Kowalski}, \citenamefont {Scholes}, \citenamefont {Yuen-Zhou}, \citenamefont
		{Baldo}, \citenamefont {Freedman}, \citenamefont {Goldsmith}, \citenamefont
		{Goodson} \emph {et~al.}}]{wasielewski2020exploiting}%
	\BibitemOpen
	\bibfield  {author} {\bibinfo {author} {\bibfnamefont {M.~R.}\ \bibnamefont
			{Wasielewski}}, \bibinfo {author} {\bibfnamefont {M.~D.}\ \bibnamefont
			{Forbes}}, \bibinfo {author} {\bibfnamefont {N.~L.}\ \bibnamefont {Frank}},
		\bibinfo {author} {\bibfnamefont {K.}~\bibnamefont {Kowalski}}, \bibinfo
		{author} {\bibfnamefont {G.~D.}\ \bibnamefont {Scholes}}, \bibinfo {author}
		{\bibfnamefont {J.}~\bibnamefont {Yuen-Zhou}}, \bibinfo {author}
		{\bibfnamefont {M.~A.}\ \bibnamefont {Baldo}}, \bibinfo {author}
		{\bibfnamefont {D.~E.}\ \bibnamefont {Freedman}}, \bibinfo {author}
		{\bibfnamefont {R.~H.}\ \bibnamefont {Goldsmith}}, \bibinfo {author}
		{\bibfnamefont {T.}~\bibnamefont {Goodson}},  \emph {et~al.},\ }\href
	{\doibase 10.1038/s41570-020-0200-5} {\bibfield  {journal} {\bibinfo
			{journal} {Nature Reviews Chemistry}, \bibinfo {pages} {1}} (\bibinfo
		{year} {2020})}\BibitemShut {NoStop}%
	\bibitem [{\citenamefont {Francica}\ \emph {et~al.}(2017)\citenamefont
		{Francica}, \citenamefont {Goold}, \citenamefont {Plastina},\ and\
		\citenamefont {Paternostro}}]{Gianluca:17}%
	\BibitemOpen
	\bibfield  {author} {\bibinfo {author} {\bibfnamefont {G.}~\bibnamefont
			{Francica}}, \bibinfo {author} {\bibfnamefont {J.}~\bibnamefont {Goold}},
		\bibinfo {author} {\bibfnamefont {F.}~\bibnamefont {Plastina}}, \ and\
		\bibinfo {author} {\bibfnamefont {M.}~\bibnamefont {Paternostro}},\ }\href
	{\doibase 10.1038/s41534-017-0012-8} {\bibfield  {journal} {\bibinfo
			{journal} {npj Quantum Information}\ }\textbf {\bibinfo {volume} {1}},\
		\bibinfo {pages} {12} (\bibinfo {year} {2017})}\BibitemShut {NoStop}%
	\bibitem [{\citenamefont {Obando}\ \emph {et~al.}(2015)\citenamefont {Obando},
		\citenamefont {Paula},\ and\ \citenamefont {Sarandy}}]{Obando:15}%
	\BibitemOpen
	\bibfield  {author} {\bibinfo {author} {\bibfnamefont {P.~C.}\ \bibnamefont
			{Obando}}, \bibinfo {author} {\bibfnamefont {F.~M.}\ \bibnamefont {Paula}}, \
		and\ \bibinfo {author} {\bibfnamefont {M.~S.}\ \bibnamefont {Sarandy}},\
	}\href {\doibase 10.1103/PhysRevA.92.032307} {\bibfield  {journal} {\bibinfo
			{journal} {Phys. Rev. A}\ }\textbf {\bibinfo {volume} {92}},\ \bibinfo
		{pages} {032307} (\bibinfo {year} {2015})}\BibitemShut {NoStop}%
	\bibitem [{\citenamefont {Ollivier}\ and\ \citenamefont
		{Zurek}(2001)}]{Zurek:01}%
	\BibitemOpen
	\bibfield  {author} {\bibinfo {author} {\bibfnamefont {H.}~\bibnamefont
			{Ollivier}}\ and\ \bibinfo {author} {\bibfnamefont {W.~H.}\ \bibnamefont
			{Zurek}},\ }\href {\doibase 10.1103/PhysRevLett.88.017901} {\bibfield
		{journal} {\bibinfo  {journal} {Phys. Rev. Lett.}\ }\textbf {\bibinfo
			{volume} {88}},\ \bibinfo {pages} {017901} (\bibinfo {year}
		{2001})}\BibitemShut {NoStop}%
	\bibitem [{\citenamefont {Momma}\ and\ \citenamefont
		{Izumi}(2011)}]{momma2011vesta}%
	\BibitemOpen
	\bibfield  {author} {\bibinfo {author} {\bibfnamefont {K.}~\bibnamefont
			{Momma}}\ and\ \bibinfo {author} {\bibfnamefont {F.}~\bibnamefont {Izumi}},\
	}\href {\doibase 10.1107/S0021889811038970} {\bibfield  {journal} {\bibinfo
			{journal} {Journal of applied crystallography}\ }\textbf {\bibinfo {volume}
			{44}},\ \bibinfo {pages} {1272} (\bibinfo {year} {2011})}\BibitemShut
	{NoStop}%
\end{thebibliography}
\end{document}